\documentstyle[12pt,aaspp4,amssym]{article}

\received{January 23 1998}
\accepted{May 18 1998}
\slugcomment{Submitted to the Astrophysical Journal.}

\lefthead{M. Georganopoulos \& A.P. Marscher}
\righthead{The Blazar Inner Jet}

\begin{document}

\title{A Viewing Angle - Kinetic Luminosity Unification \\
 Scheme For BL Lacertae Objects.}

\author{Markos Georganopoulos and Alan P. Marscher}
\affil{Department of Astronomy , Boston University, 725 Commonwealth Avenue, Boston,  MA 02215}
%\authoraddr{}
%\authoremail{}

\begin{abstract}
 We propose a  unified classification for BL Lac
objects (BLs), focusing on the synchrotron peak frequency $\nu_{s}$ 
of the spectral energy distribution. The  unification scheme is 
based on the angle $\Theta$ that describes  the orientation of the 
relativistic jet and on the electron kinetic luminosity $\Lambda_{kin}$ of the jet. We 
assume that $\Lambda_{kin}$ scales with the
size of the jet $r$ in a self-similar fashion ($\Lambda_{kin}\propto r^2 $),
 as supported by  observational data. 
The jets are self-similar in geometry and have the same pressure and median
magnetic field at the inlet, independent of size. The self-similarity is
 broken for the highest energy electrons, which radiate mainly at high 
frequencies, since for large sources they suffer more severe radiative
energy losses over a given fraction of the jet length.
We calculate the optically thin synchrotron spectrum using an accelerating 
inner jet model based on simple relativistic gas dynamics
and show that it can fit the observed infrared to X-ray spectrum of
PKS 2155--304.
We couple the accelerating jet model to the unification scheme and 
compare the results  to complete samples of 
 BLs. The  negative apparent evolution of  X-ray 
selected BLs  is explained as a result of  positive evolution
of the jet electron kinetic luminosity $\Lambda_{kin}$. We review observational
 arguments  in 
favor of the existence of  scaled-down accretion disks and 
broad emission-line regions in BLs. The proposed unification 
scheme can  explain the lack of observed broad emission lines 
in X-ray selected 
BLs, as well as the existence of those lines preferentially in
luminous radio-selected BLs. Finally, we review observational arguments
that suggest the extension of this  unification scheme to 
all blazars.

\end{abstract}

\keywords{radiation mechanisms: non-thermal --- BL Lacertae objects: general 
--- galaxies: jets}

\section{INTRODUCTION}

The family of blazars includes those AGN  (active galactic nuclei)
that are characterized by compact 
radio morphology and  variable, polarized nonthermal continuum.
Apparent superluminal motion and $\gamma$-ray emission are also  common 
properties among 
blazars  (for a review of their properties see Urry \& Padovani 1995).
When the spectrum exhibits the usual quasar-like broad emission 
lines, the 
object is classified as a  flat spectrum  radio quasar (FSRQ). 
On the other hand, the term
BL Lacertae  object (BL) is reserved for the lineless or almost
lineless blazars. The parent population   of  FSRQs (\cite  {padovani92}) 
and BLs (\cite{padovani90}) is thought 
to consist of  Fanaroff-Riley type II  (FR II) 
and Fanaroff-Riley type I  (FR I) radio galaxies (\cite{fanaroff74}),
respectively.

The standard interpretation of the nonthermal continuum radiation of blazars 
 from radio to $\gamma$ rays is  synchrotron 
and inverse Compton emission from a collimated relativistic plasma jet 
(\cite{blandford78}).
The currently popular paradigm involves the existence and dynamical 
consequences  of a magnetic field that
threads an  accretion disk around a massive black hole (\cite{blandford82}).
 The collimation and 
acceleration of the jet in such a scenario is currently a field of active 
research;
both analytical (e.g. \cite{appl96}) and numerical (e.g. \cite{ouyed97}) work 
 shows promising results thus far.
 
The observational taxonomy of BLs consists, according to the method 
of discovery,
of  radio selected (RBLs) and X-ray 
selected (XBLs) sources. According to Padovani \& Giommi (1995), this
  corresponds  to a 
more physical dichotomy, based on the frequency at which the peak 
of the synchrotron spectral energy distribution (SED, quantified by 
$\nu L_{\nu}$) occurs: 
low-frequency peaked BLs (LBLs), which are mostly RBLs,
and high-frequency peaked BL Lac objects (HBLs), which are mostly XBLs. 
The XBLs  are less variable and polarized than RBLs 
(\cite{jannuzi94}) and
 have flatter optical spectra (\cite{ledden85}). The RBLs are more
core-dominated (\cite{perlman93}) and for a given X-ray luminosity are 
more luminous at radio
and optical frequencies than XBLs (\cite{maraschi86}). RBLs are
characterized by a weak positive evolution 
(i.e., dependence of number counts on redshift; \cite{stickel91}), 
while XBLs 
exhibit a negative evolution (\cite{perlman96}). The recent discovery of a 
population  of intermediate BLs (\cite{laurent97};  \cite{perlman98})
suggests a continuous family of 
objects rather than two separate classes.

Maraschi et al. (1986) proposed that the X-ray emission comes from a compact, 
less  beamed region, while the radio emission comes from an extended, more 
beamed region. According to  this  ``orientation hypothesis'',
  the RBLs 
form a small angle ($\Theta \la 10^{\circ}$) between the line of sight 
 and the    
 direction of motion of the emitting plasma, 
while the XBLs form a larger angle ($10^{\circ} \la \Theta \la 30^{\circ}$).
A more recent interpretation of the differences between XBLs and RBLs 
(\cite{padovani95}) is based on the cutoff frequency of the synchrotron 
SED.
According to this ``SED-cutoff hypothesis,''  most  of the BLs are 
characterized by a cutoff in the IR/optical band, and these are the
radio selected objects. The small fraction of BLs that have a 
cutoff at UV/X-ray energies are the X-ray selected BL Lac objects.
This model does not offer a physical explanation for the different cutoff 
frequencies  of the SED.

There is an ongoing  debate in the literature about the actual reason for the 
differences between XBLs and RBLs. Padovani \& Giommi (1995) argue that the
observed range of the core dominance parameter R (the ratio of  core to
extended radio flux) of the XBLs does not agree with the  
orientation hypothesis. 
On the other hand, Kollgaard et al. (1996a) conclude  that the fact that the 
mean core
radio power of RBLs exceeds that of XBLs by more  than an order of magnitude
for sources selected according to extended radio power, is incompatible with the 
SED-cutoff hypothesis. Comparison of VLBI images of RBLs and XBLs
 (\cite{kollgaard96}) suggests that  the jets in XBLs fade faster
 than in 
RBLs, which supports the idea that factors other than orientation 
are also important in explaining the differences between XBLs and RBLs. 

Urry \& Padovani (1995) note that the  SED-cutoff hypothesis cannot 
explain the different polarization properties of XBLs and RBLs, 
particularly
the stability of the polarization angle in the XBLs. Lamer, Brunner, 
\& Staubert (1996)  analyze a sample of BL Lac objects observed with the 
PSPC
detector on board the ROSAT satellite and note that the SED-cutoff 
hypothesis can explain the range of the crossover frequency between the 
soft and the hard components in the X-ray regime  better than the 
orientation hypothesis. Sambruna, Maraschi, \& Urry (1996), using complete
samples of BLs, reach the conclusion that the range of the peak
frequency of the synchrotron energy distribution in BLs  cannot 
be reproduced under the  orientation hypothesis.
An alternative scenario, which connects
the synchrotron peak luminosity $L_{s}$ to
the synchrotron peak frequency $\nu_{s}$ of the SED of the BLs 
through an empirical 
relation, has been proposed 
recently  by Fossati et al. (1997).  
One feature that this scenario reproduces more successfully  than the 
previous two is the redshift distribution of XBLs and RBLs.

In this work we propose that two parameters determine the observed 
characteristics  of a BL. The first which must play a role, 
given the relativistic nature of 
the jet flow, is the angle $\Theta$ formed between the line of sight and the jet axis. The second  is the electron kinetic luminosity $\Lambda_{kin}$ of the jet.
We start by developing and testing a numerical code for the accelerating 
inner jet model, which we  use  as our working hypothesis
for the physical description of the jet.
Invoking  recent observational studies, we propose a  
new classification scheme,
 whose main observational parameter is  the synchrotron peak frequency 
$\nu_{s}$ of the 
SED of BLs. 
We then introduce  the theoretical $\Theta$--$\Lambda$  unification 
for BLs, based on two parameters: 
the angle $\Theta$ between the jet axis and the line of sight, and  the electron kinetic
luminosity $\Lambda_{kin}$ of the jet. These are
 coupled to  a simple self-similar 
jet description  that relates  $\Lambda_{kin}$ to the size of the jet 
($\Lambda_{kin} \propto r^2$). 
We use the accelerating jet model to 
show how this unification scheme can be used to explain several characteristics
of BLs, including the negative apparent evolution of the XBLs. We also
show, using  the self-similarity scenario we propose, 
that a picture in which BLs have a scaled-down accretion disk
and broad emission-line region (BELR) environment similar to those observed
 in FSRQs and high 
polarization
quasars (HPQs), is in agreement with observations.
(We will use the terms FSRQ and HPQ interchangeably to signify all the 
non-BL  blazars).
Finally, we review observational evidence that supports 
a unified picture for all 
blazars under the above self-similar scheme.

\section{THE JET MODEL\label{model}}

A first  approach in calculating the synchrotron emission  produced by an
accelerating and collimating plasma flow  was presented by Marscher (1980): 
In  the accelerating inner jet model 
(based on the work of \cite{blandford74} and \cite{reynolds82}), continuously generated
ultrarelativistic plasma
is confined by  pressure (hydrostatic and/or magnetohydrodynamic) that 
decreases
along the jet axis. The internal energy of the 
plasma is converted to bulk kinetic energy and the jet is accelerated and 
focused. The electrons
interact with a predominately random magnetic field and cool through
synchrotron radiation and adiabatic expansion. At the
same time, inverse Compton losses become important when the
synchrotron photon energy density becomes comparable to the magnetic field
energy density. Higher frequency radiation is  emitted close to
 the base of the jet, since only there are the electron energies
high enough to do so. Lower frequency photons are emitted
there as well as farther downstream. The velocity of
the jet increases with distance; this implies larger Doppler boosting
for greater distances down the jet  and therefore at lower
frequencies, out to the point where the Lorentz
factor $\Gamma \la \Theta^{-1} $. The relevance of this approach to the BL Lac phenomenon
was strengthened through studies that suggested that  the X-ray emitting
region is less relativistically boosted than that of the radio emission
 (\cite {maraschi86}; \cite{urry91}). 

We use the following  phenomenological description  for the accelerating 
inner jet:
ultrarelativistic plasma (mean ratio of total to 
rest mass energy  per particle  $\gamma \gg 1$ in the proper frame of the fluid)
is continuously injected at the base of the jet.
The bulk flow of the plasma at the injection point is parameterized by the
bulk  Lorentz factor $\Gamma_{\star}$. 
The axially symmetric 
pressure profile, with the pressure on the symmetry  axis  being less than 
the equatorial pressure,
 defines a preferred direction 
for the expansion of the plasma. The plasma flow is progressively accelerated 
and
collimated, converting its internal  energy to bulk kinetic energy.
The adopted phenomenological description of ultrarelativistic particle 
injection at
the base of the jet can be linked, for example,  to particle acceleration 
in a standing shock front
in the plasma flow, formed due to a drop in the confining pressure (\cite{gomez97}).
 Since the emission at  higher frequencies is confined to a region 
closer to the base of the jet  than at lower frequencies, we expect 
(\cite{marscher80}; \cite{ghisellini89}) shorter
variability timescales at higher frequencies, given that the acceleration
of the flow is mild.
In addition, the Doppler effect is  comparatively more important for the 
lower than for the higher 
frequencies. Therefore the jet orientation affects the lower frequencies
more than the higher ones.

\subsection{Flow description}

We adopt a flow description (see appendix \ref{gas}) similar to that of 
Blandford \& Rees (1974) and Marscher (1980), the latter of which
 considered as the base of the jet
the sonic point, where the bulk Lorentz factor has the value 
$\Gamma_{\star}=\sqrt{3/2}$. 
If we identify the base of the jet instead  with a standing shock,  we can 
relax the assumption  
that the 
flow is sonic at the base of the jet, and allow for higher values of the bulk
Lorentz factor $\Gamma_{\star}$ at that point. The minimum distance of the
base of the jet from the accretion disk is obtained by requiring that the 
electron Thomson losses due to the accretion disk photons are less significant than the synchrotron losses (see Appendix \ref{thomson}).  The quantities that 
describe the flow
are the total (electron + proton) kinetic luminosity  $\Lambda_{t}$ of the 
jet, the radius $r_{\star}$ of the base of the
jet,  and  the exponent $\epsilon$ and the length scale $z_{\star}$ that 
describe how fast the jet opens and
accelerates. 
Assuming  energy equipartition between electrons and protons, the electron 
kinetic luminosity  $\Lambda_{kin}$ is 
\begin{equation}
 \Lambda_{kin}=\Lambda_{t}/2.
\end{equation}
Thus  half of the injected energy is in protons that do not
radiate, and therefore we can reasonably approximate the flow as adiabatic.

The bulk Lorentz factor as a function of distance along
the jet axis is given by equation  (\ref{eq:Gamma}):
\begin{equation}
\Gamma(z)=\Gamma_{\star}\left(\frac{z}{z_{\star}}\right)^{\epsilon},
\end{equation} 
while the radius of the jet is given by a modified form of equation (\ref{eq:radius}): 
\begin{equation}
r(z)=r_{\star}\left(\frac{z}{z_{\star}}\right)^
{\frac{3\epsilon}{2}}   (\Gamma_{\star}^{2}-1) ^{\frac{1}{4}}     \left[\Gamma_{\star}^{2}\left(\frac{z}{z_{\star}}\right)^{2\epsilon}-
1\right] ^{-\frac{1}{4}}. 
\label{eq:realradius} 
\end{equation}
For 3-D steady flow, this equation applies to each streamline, but with 
$r(z)$ replaced by $r_{str}$. The flow velocity is tangent to the
streamlines and the loci of constant scalar quantities such as density 
are surfaces that are perpendicular to the streamlines.
Note that for $\Gamma_{\star}=\sqrt{3/2}$ equation (\ref{eq:realradius})
 is reduced
 to equation (\ref{eq:radius}).

\subsection{Magnetic field}

For the purpose of this work, we will consider the  magnetic field 
to be mainly random in direction, with 
a small ordered component parallel to the jet axis. Under the assumption of
an isotropic electron  
distribution (see \S\S \ref{eed}), the  choice of magnetic
field geometry is not critical, since we are interested here  in the total 
and not
the polarized flux. Using simple flux conservation arguments,
 we assume that the
 magnetic field in the fluid  frame decays as $B\propto r^{-m},
\hspace{.1in} 1\leq m \leq 2$, or in terms of the jet axial coordinate $z$:
\begin{eqnarray}
B(z)=B_{\star}\left(\frac{z}{z_{\star}}\right)^
{\frac{-3m\epsilon}{2}}   (\Gamma_{\star}^{2}-1) ^{-\frac{m}{4}}  \nonumber \\[0.25cm]\times   \left[\Gamma_{\star}^{2}\left(\frac{z}{z_{\star}}\right)^{2\epsilon}-
1\right] ^{\frac{m}{4}}.
\label{eq:field}
\end{eqnarray}

\subsection{Electron energy losses \label{losses}}

We assume that the electrons are injected at the base of the jet and 
propagate downstream while losing energy due to  adiabatic and synchrotron 
energy 
losses, without any  reacceleration (except at downstream shocks fronts, the
radiation from which is not considered here).
 The electrons will also 
suffer inverse Compton losses from the 
synchrotron photons they produce (synchrotron self Compton (SSC) model, e.g., 
\cite{marscher94}) as well as 
from photons coming either from
the broad-line clouds (\cite{sikora94}) or from the putative accretion disk
(\cite{dermer93}) (external Compton (EC)). In our treatment we require that the base of the jet
is sufficiently far from the central engine, so that the electron 
Thomson losses due  to scattering of the accretion disk radiation 
are smaller than the synchrotron
losses (see Appendix B). 
 A self-consistent treatment of inverse 
Compton scattering is extremely complicated, 
since the local photon energy density results from photons that have been 
emitted in different regions of the jet at different retarded times. 
Observations using the Compton Gamma Ray Observatory (CGRO) (\cite{montigny95}; \cite{dondi95}; \cite{fossati98})
 suggest that during flares  the Compton  luminosity is usually 
greater than the synchrotron   in  FSRQs by  $\approx $ 1--2 orders 
of magnitude, but is  
comparable to, or less than,   the synchrotron luminosity in BLs, 
and during  low states
of $\gamma$-ray activity the synchrotron luminosity  dominates the power output
 in BLs. In SSC models the beaming cone of the synchrotron and Compton radiation
is the same, and the ratio of apparent  luminosities  of the two components 
equals the ratio  of synchrotron to Compton
electron losses. 

In EC models the beaming cone of the inverse Compton
radiation is more narrow than the synchrotron cone (\cite{dermer95}). 
Statistically, in the case of EC models, even if
Compton losses are equal  to the synchrotron losses, in any BL sample
selected  at frequencies  dominated by synchrotron emission, 
one would expect that
 the Compton luminosity would be significantly greater
than the synchrotron one for a fraction of the sources equal to the
ratio of the beaming cone solid angles. Since this is not observed,
it seems plausible to assume that the Compton losses even in EC dominated
BL models are not the main energy loss mechanism for BLs.

We combine the adiabatic and synchrotron losses to obtain the electron
energy-loss equation in the fluid frame:

\begin{eqnarray}
\frac{d\gamma(z)}{dz}&=&\left(\frac{d\gamma(z)}{dz}\right)_{synch}+   
                \left(\frac{d\gamma(z)}{dz}\right)_{ad} 
 \nonumber \\[0.25cm]&=&
-C_1\gamma^2\left(\frac{z}{z_{\star}}\right)^{-3m\epsilon} 
  (\Gamma_{\star}^{2}-1) ^{-\frac{m}{2}}  
\nonumber \\[0.25cm]& &
\times   \left[\Gamma_{\star}^{2}\left(\frac{z}{z_{\star}}\right)^{2\epsilon}-
1\right] ^{\frac{m-1}{2}}-\frac{\epsilon \gamma}{z},
\label{eq:losses}
\end{eqnarray}

where
\begin{equation}
C_1=(\frac{2}{3})^2\frac{e^4}{m^3c^6} B_{\star}^2. \nonumber
\end{equation}

 Since  $(d\gamma/dz)_{synch}\propto\gamma^2$ while
$(d\gamma/dz)_{ad}\propto\gamma$,  synchrotron losses dominate for
 the higher energy electrons.  The Lorentz factor $\gamma$  
describes the internal kinetic energy of the  electrons in the fluid proper frame. Solving  for the
 electron energy, we obtain

\begin{equation}
\gamma(z)=\left(\frac{z}{z_{\star}}\right)^{-\epsilon}
\left[\frac{1}{\gamma(z_{\star})} + C_1  (\Gamma_{\star}^{2}-1) ^{-\frac{m}{2}}      I(z)\right]^{-1},
\label{eq:gamma}
\end{equation}
where the integral
\begin{equation}
 I(z)=\int_{z_{\star}}^{z}
 \left(\frac{z}{z_{\star}}\right)^{-(3m+1)\epsilon}
 \left(\Gamma_{\star}^2\left(\frac{z}{z_{\star}}\right) ^{2\epsilon}-1\right)
 ^{\frac{m-1}{2}} dz 
\end{equation}
must  be solved numerically for $m \neq 1$.

\subsection{Electron energy distribution \label{eed}}

The EED is, in principle,
 determined by the
particle acceleration  and the dominant energy loss mechanisms acting
 simultaneously on the electrons (\cite{kirk97}).
 These two factors should determine 
the shape   and the cutoffs of the 
distribution. In our model we assume that the initially injected
EED has a power law form:
\begin{equation}
N(\gamma_{\star},z_{\star})=N_{\star} \gamma_{\star}^{-s},\hspace
 {.05in} \gamma_{\star} \in 
 [\gamma_{min}(z_{\star}),\gamma_{max}(z_{\star})].
\end{equation}
We further assume  that the initial electron energy distribution (EED)
 remains isotropic in the comoving frame,
through redistribution of the electron pitch angles due to their scattering
on plasma waves. The EED distribution is therefore a function
of energy and position along the jet axis only.
The normalization factor $N_{\star}$ is related to the comoving energy 
density $e_{el\star}$ at the base of the jet
\begin{equation}
e_{el\star}=\int_{\gamma_{min}}^{\gamma_{max}}\gamma N_{\star}\gamma^{-s}d\gamma=
\frac{N_{\star}}{2-s}(\gamma_{max}^{2-s}-\gamma_{min}^{2-s}).
\end{equation}
Using equation (\ref{eq:lambda}) for the base of the jet 
($A=r_{\star}^2$) and the relation  $e_{\star}=(3/4)w_{\star}$ 
(see Appendix \ref{gas}), together with
the assumed energy equipartition between electrons and protons 
($e_{el\star}=e_{\star}/2$), we obtain

\begin{equation}
N_{\star}=\frac{3(2-s)\Lambda_{kin}}{4\Gamma_{\star}^2\beta_{\star} c(\gamma_{max}^{2-s}-\gamma_{min}^{2-s}).}
\end{equation}

In our treatment it is implicitly assumed that the acceleration time is much
shorter than the radiative loss time for the considered frequencies, and we can
 therefore separate  the region where acceleration
takes place from the radiative zone.
This assumption  may break down for the  highest synchrotron frequencies, 
although even in extreme objects like Mkn 421 the 1--10 keV variability
behavior indicates  that this assumption is valid (\cite{kirk97}). 
We therefore consider this approach 
adequate for synchrotron
radiation  with photon energy up to  $\sim 10 $ keV.
Conservation of particles and eq. (\ref{eq:gamma}) give the  evolved energy
 distribution:
\begin{eqnarray}
N(\gamma,z)=N_{\star}\gamma^{-s}\left(\frac{z}{z_{\star}}\right)^
{-\epsilon(s+2)}\times
 \nonumber \\[0.25cm]
 \left(1- \gamma \left(\frac{z}{z_{\star}}\right)^{\epsilon} C_1 (\Gamma_{\star}^{2}-1) ^{-\frac{m}{2}} I(z)
 \right)^{s-2}.
\label{eq:dens}
\end{eqnarray}

This profile is no longer a power law, and on  a log-log plot it 
gradually steepens  toward  higher frequencies. Note also that
the EED evolves as the plasma moves downstream. 
This  gives rise to  a non-power law locally emitted  synchrotron spectrum 
in the local flow frame, 
which  changes with $z$ in frequency cutoff and shape.

\subsection{Relativistic transformations}
Given the magnetic field and the EED in the local fluid proper frame we can calculate the synchrotron  emission
coefficient in the fluid frame using standard synchrotron formulae
(\cite{pacholczyk70}):
\begin{eqnarray}
\epsilon_\nu(z)=\frac{\sqrt{3} e^3}{4\pi mc^2}B(z)\sin\phi'  
 \nonumber \\[0.25cm] \times                     
     \int_{\gamma_{min}(z)}^{\gamma_{max}(z)}N(\gamma,z)F(x)d\gamma,
\label{eq:emcoef}
\end{eqnarray}
where 
\begin{equation}
F(x)=x\int_x^{\infty}K_{5/3}(z)dz.
\end{equation}
$K_{5/3}$ is the modified Bessel function of order $ 5/3 $ and $x$ a
 dimensionless frequency:
\begin{eqnarray}
x&=&\frac{\nu}{\nu_{crit}(z)}, \nonumber 
\end{eqnarray}
where
\begin{eqnarray} 
 \nu_{crit}(z)&=&\frac{3e}{4\pi mc}B(z)\sin\phi'\gamma(z)^2. \nonumber
\end{eqnarray}
In eq. (\ref{eq:emcoef}) $\phi'$   is the angle between the magnetic field and the line of sight in the fluid  frame. In order to calculate this angle we first 
need to transform the angle $\theta$ between the line of sight and the local flow 
velocity from the observer's frame to the fluid frame using standard relativistic
 aberration formulae (\cite{rl79}):

\begin{equation}
\cos\theta'=\frac{\cos\theta-\beta}{1-\beta \cos\theta}.
\end{equation}

Having now the angle $\theta'$ between the line of sight and the local flow velocity
 in the local fluid  frame, we can obtain  $\phi'$ from   the angle
between the magnetic field and the local flow velocity in the fluid proper 
frame.
 
 To transform the emission coefficient to the
observer's (unprimed) frame we need to take into account the transformations
between this frame and the fluid (primed) frame. If $z$ is
the redshift and $\delta=1/(\gamma(1-\beta \cos\;\theta))$ is the Doppler
boosting factor, where $\theta$ is the angle between the sight and the fluid velocity,  then the frequency in the observer's  frame is
$\nu=\nu' \delta/(1+z) $ and the emission coefficient is (\cite{rl79}):
\begin{equation}
\epsilon_{\nu}=\left(\frac{\delta}{1+z}\right)^2\epsilon_{\nu'}'.
\end{equation}
Having now the synchrotron emission coefficient in the observer's
 frame, we can 
perform the radiative transfer and calculate the total 
synchrotron flux for optically thin frequencies.

\section{MODEL SPECTRAL ENERGY DISTRIBUTIONS\label{results}}

\subsection{The parameter space\label{parspace}}
\begin{figure}
\epsscale{0.8}
\plotone{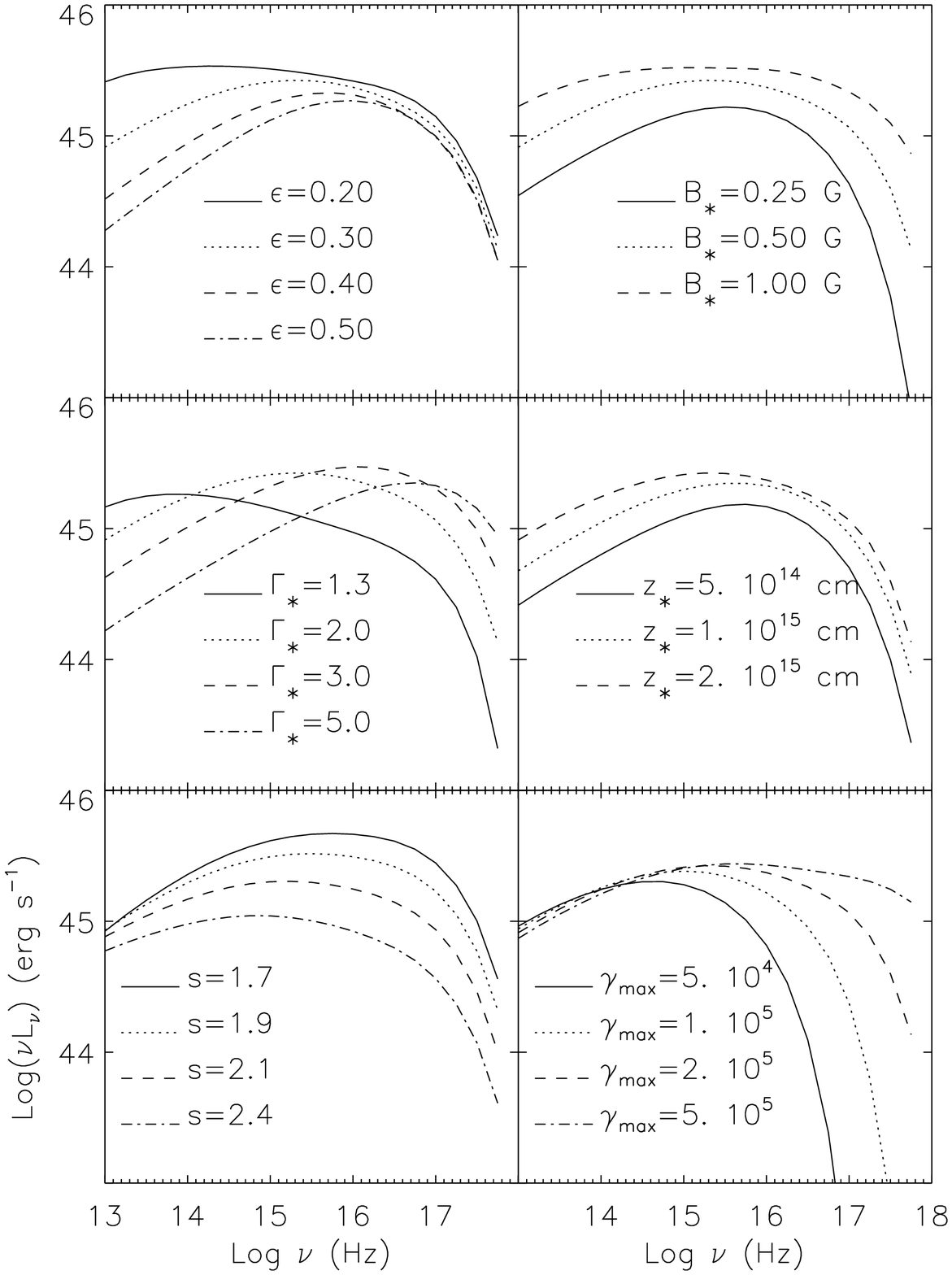}
\caption{ The model spectral energy distribution (SED), for values of 
 the parameters as given in Table \ref{tbl1}, except for the single
parameter allowed to vary in each panel. Here, $L_{\nu}$ is the apparent
luminosity per unit frequency.}
\label{parameters}
\end{figure}

We have performed an extensive study of the model parameter space in order to  
examine the response of the resultant  SED to  a variation of each
physical parameter about the fiducial values given   in Table \ref{tbl1}.
 The meaning of the symbols is the same as in \S
\ref{model}. 
In Figure \ref{parameters} we allow one of the model parameters to vary in
each panel and  plot the resulting SED. 

As $\epsilon$ increases 
(corresponding to a more poorly collimated jet)
 the peak apparent synchrotron luminosity $L_{s}$ decreases.
This is due to an increase of the adiabatic losses and a decrease
of the synchrotron power (cf. eq. \ref{eq:losses}), 
 regardless of the value of $m$.
At the same time  the peak frequency $\nu_{s}$ of the SED 
increases, since the synchrotron losses decrease, and the effective
break of the EED shifts to higher energies.
An increase in the magnetic field amounts to an increase in $L_{s}$ and a 
decrease in $\nu_{s}$. This is expected since a higher magnetic field  
results in  higher synchrotron power and hence losses, primarily
 for the higher energy
electrons. 

The behavior of the SED as we increase $\Gamma_{\star}$ is less 
straightforward. As $\Gamma_{\star}$ increases, $\nu_{s}$ increases, the
IR luminosity decreases, the X-ray luminosity increases, and  $L_{s}$
increases until $\Gamma_{\star}$  exceeds a critical value,
 beyond which it decreases.
 The key in understanding
this behavior is that 
 the jet emission is stratified, with the higher frequencies, which emerge 
closer to the base of the jet, being characterized by smaller Lorentz factors.
As we increase  $\Gamma_{\star}$, the effective  Lorentz factor $\Gamma_{\nu}$
for each frequency
increases, and the angle ($\approx 1/\Gamma_{\nu}$) 
into which the radiation is beamed decreases. In the IR regime 
(for the parameters chosen in Table \ref{tbl1})  an increase in
$\Gamma_{\star}$ beams most of the radiation into  a cone with an opening angle
smaller than $\Theta$; this causes the Doppler de-amplification we observe.
In X-rays the increase in $\Gamma_{\star}$ beams the radiation into a 
narrower cone  that still includes the line of sight and therefore the observed luminosity is more 
 Doppler boosted.  This frequency-dependent
 response causes $\nu_{s}$ to shift
toward higher frequencies. $L_{s}$ increases as long as the Doppler 
boosting of the higher frequencies can compensate for the Doppler 
de-amplification at the lower frequencies. After a certain point 
(for the parameter values we use, this occurs  at $\Gamma_{\star} \approx 3$),
 the observed 
peak luminosity  declines, since most of the emitted power is beamed
into a cone that does not include our line of sight.

An increase in $z_{\star}$ produces a behavior similar to that exhibited by 
the increase in $B_{\star}$. This is expected, since an increase in $z_{\star}$
causes the  magnetic field to decline more slowly with time in the electron's 
frame. That is, the electron spends more time in the roughly uniform magnetic
field region as the size of that region increases.

An increase in $s$ (a steeper EED) produces
a steeper spectrum, as expected. As the EED 
becomes steeper, the energy content of the high energy tail of the EED 
decreases. Since these are the electrons that lose energy faster 
and emit at  higher frequencies, both  $L_{s}$ and $\nu_{s}$ decrease as
the EED becomes steeper.
An increase in the upper cutoff $\gamma_{max}$ of the EED affects 
dramatically the high frequency tail of the SED, making the X-ray spectrum
flatter and increasing both  $L_{s}$ and $\nu_{s}$. The effect on the 
low frequencies is minimal, as expected, since the high energy electrons
radiate a small fraction of their power there.
The behavior of the SED
resembles that described by  Padovani \& Giommi (1995), 
who attribute  the cause of the
differences between XBLs and RBLs to different  values of $\gamma_{max}$. 

The exponent $m$, 
which describes the  evolution of the comoving magnetic field along the jet 
axis,
and the radius of the jet $r_{\star}$ (not shown in Figure \ref{parameters}) 
only weakly affect the SED.
Increasing m from m=1 to m=2 decreases  the IR luminosity by a factor of
$\sim 2 $, while the X-ray luminosity remains essentially the same.
Increasing $r_{\star}$ by a factor of four reduces the X-ray luminosity 
by a factor of $ \sim 2 $, while the IR luminosity remains constant.
Note, however, that 
in a time dependent simulation the minimum variability
timescale [$ t_{var} \gtrsim r_{\star}\sin \Theta/ c$] 
provides an upper limit to $r_{\star}$. 
The value of the lower cutoff $\gamma_{min}$ of the EED should in principle
be derived from   the particle acceleration theory. 
 For a given jet electron kinetic luminosity, as $\gamma_{min}$ increases,
 the particle
density and the observed synchrotron luminosity increase,
since the average energy per particle increases. 
Hence, $\Lambda_{kin}$ and $\gamma_{min}$ are closely connected and only one
($\Lambda_{kin}$) needs to be treated as a free parameter unless the value
of the density is needed.
For the purpose of this
work we will assign  
the  values $m$=1 and  $\gamma_{min}=10^{2}$. 

\begin{figure}
\epsscale{0.8}
\plotone{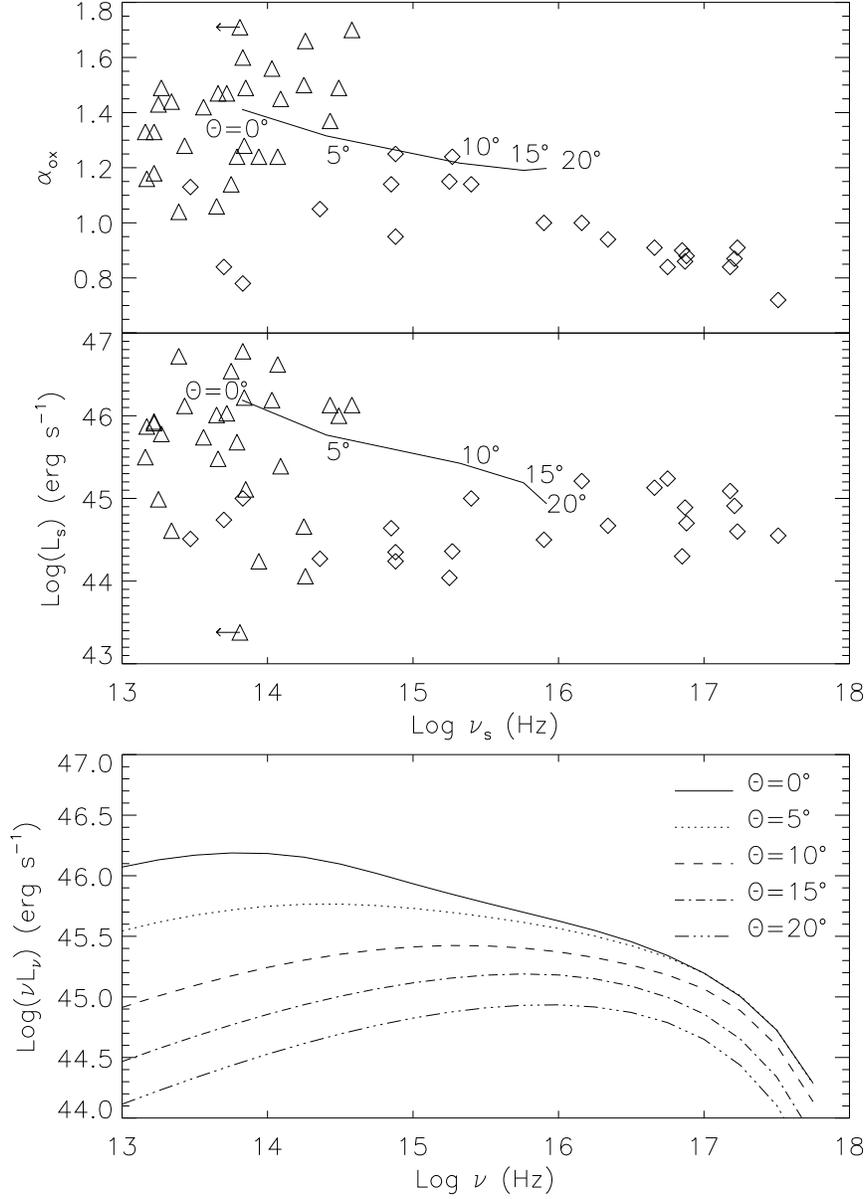}
\caption{ The (bottom panel) SED for values of the parameters given in 
Table \ref{tbl1}, for various  observing angles $\Theta$ between 
the jet axis and the line of sight. Peak luminosity $L_{s}$ (middle panel) and 
optical to X-ray spectral index $\alpha_{ox}$ (top panel)  versus 
observed peak synchrotron frequency $\nu_{s}$. The data points  correspond 
to samples of RBLs (triangles) and XBLs (diamonds) (\protect\cite{sambruna96}), and 
the curves correspond to the  model.}
\label{angle}
\end{figure}

 For a given  physical description of the jet,
the angle $\Theta$ between the jet axis and the line of sight is still
 a free parameter, in the sense that the jet can 
be oriented at any angle relative to the line of sight. 
As mentioned in the introduction,
one of the leading scenarios for the interpretation of the 
differences between RBLs and XBLs is 
based on the orientation of the jet relative to the line of sight. 
We have run our simulation using the parameters  of Table \ref{tbl1}, and 
have  allowed
the angle to vary, as  presented in Figure \ref{angle}. 
The data points refer to the
1 Jy sample of RBLs (triangles) and the \it Einstein \rm Extended Medium 
Sensitivity Survey sample of XBLs used by  Sambruna et al. (1996). 
As can be  seen in the bottom panel, as the 
angle increases from $\Theta=0^{\circ}$ to  $\Theta=20^{\circ}$ the 
synchrotron luminosity $L_{s}$ decreases by a factor of $\sim 20$ and the
synchrotron peak frequency $\nu_{s}$ increases by  $\sim  2$ orders of 
magnitude. At the same time the spectrum becomes flatter at all frequencies.
The synchrotron peak luminosity $L_{s}$ (middle panel)  
 and the optical to X-ray spectral index $a_{ox}$
 (top panel), calculated taking 
into account the luminosities at $5500$ \AA { } and at 1 keV, are plotted
versus the peak 
frequency $\nu_{s}$ of the SED.  Note that the 
model point shifts from a predominately RBL region to a predominately XBL
region in both the luminosity and spectral index plots, as the angle increases.
This behavior is not specific to the set of model parameters  used here, 
but appears in a qualitatively similar manner for a wide range of initial
conditions: As the angle increases, $L_{s}$ decreases, $\nu_{s}$ increases,
and the spectrum becomes flatter. Although these  differences 
are similar to the differences found between XBLs and RBLs, previous work
  (Sambruna et al. 1996; \cite{markos96}) has shown that a change only 
in the jet
orientation is not enough to explain the range of observed properties
of BLs, hence additional physical differences must  be 
invoked.

\subsection{Simulation of the SED of PKS 2155--304}

As an application of the model we simulate the SED of
PKS 2155--304, a relatively nearby ( redshift $Z$=0.116)  bright  XBL.  Its  $\gamma$-ray
luminosity is less than its synchrotron  (\cite{vestrand95}), which implies 
(see also
section \ref{losses})
that  synchrotron losses are more important than  inverse 
Compton losses. 
The object has been the target of two intensive multiwavelength campaigns, one 
during November 1991 (\cite{edelson95}), and one during May 1994 (\cite{urry97}).

\begin{figure}
\epsscale{0.8}
\plotone{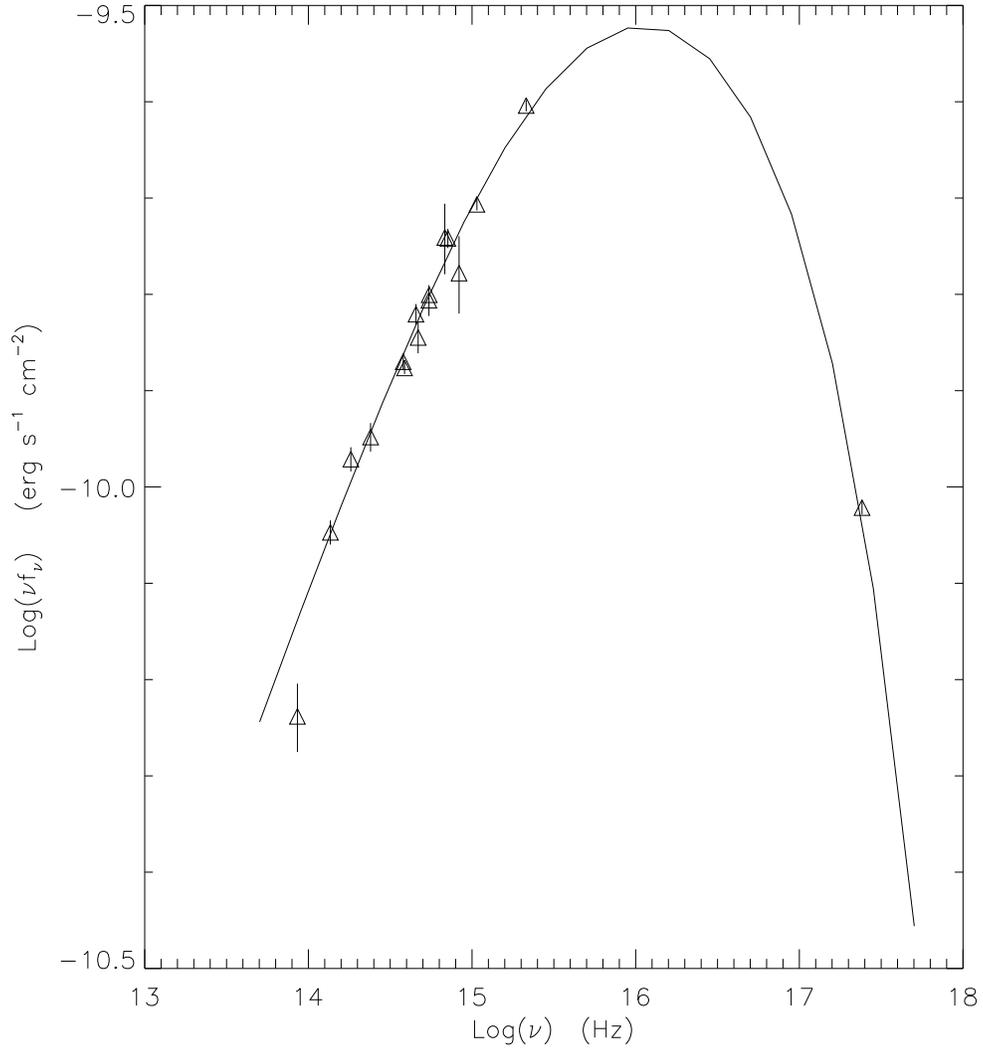}
%\plotfiddle{2155.ps}{5.in} {90.} {60.}{60.}{200.}{0.}
\caption{The spectrum of PKS 2155--304 during the multiwavelength campaign
of November 1991. The data were obtained on November 14 
during an 18--hour period. (\protect\cite{edelson95}). The solid curve
corresponds to the model fit produced using the parameters listed 
in Table \protect\ref{tbl2}.}
\label{2155}
\end{figure}

Since PKS 2155--304 is highly variable and a strictly simultaneous SED is 
practically unattainable, we must approximate a simultaneous SED with
one  constructed from fluxes measured during a time interval smaller or 
at least comparable to the timescale of the fastest significant variations.
The shortest variability timescale for the November  1991 dataset
 was $\approx0.7$ days.
We construct the SED plotted in Figure \ref{2155} using observations
in the time interval between November 14.0 and 14.76. This time interval is
approximately equal to the fastest variability timescale observed, and it 
corresponds to a local flux minimum for the optical to X-ray frequency range.
The solid curve 
in Figure \ref{2155} represents the best-fit model. The values of the 
parameters used are given in Table \ref{tbl2}. The $\chi^{2}$ 
for this fit is 49.65 for nine degrees of freedom. 
This high value can be partly attributed  to the arbitrariness
 of the
selected observed fluxes used to construct the ``base''
observed SED, but in any case is typical of fits to multiwavelength spectra
with non-complex models. 
Since an exhaustive exploration of the
parameter space is practically impossible, we cannot assert that this
is the best possible fit. To derive  a more constrained
description of  the source, we would 
need to model successfully the observed variability behavior.

\section{A NEW SED--BASED CLASSIFICATION SCHEME \label{classification}}

BL Lac objects are discovered mostly through X-ray or 
radio surveys. The existing complete samples therefore are either X-ray 
(\cite{morris91}; \cite{perlman96}) or radio flux-limited samples 
(\cite{stickel91}) and BLs are named  XBL (X-ray selected)
 or RBL (radio selected), depending  on the method of discovery. In general
RBLs have  steeper spectra than do XBLs.
The SEDs of RBLs  peak 
somewhere in the infrared and have a 
higher synchrotron peak luminosity $L_{s}$ than do the XBLs, which
 usually peak 
in the UV--soft X-ray
range (Sambruna et al. 1996). 

A recently  proposed  classification scheme for BLs 
(\cite{padovani96})
 corresponds more closely  to the observed  characteristics of the objects 
than the method of discovery. According to this
scheme BLs are classified as high frequency peak BLs 
(HBL) or low frequency peak BLs (LBL), depending on the value
of the X-ray to radio flux ratio. Objects with $f_{x}/f_{r}\lesssim 10^{-11.5}$
are classified as LBLs and objects with $f_{x}/f_{r}\gtrsim 10^{-11.5}$ 
as HBLs.
Most of the HBLs are XBLs and most of the LBLs are RBLs. 
Although this classification scheme is based on   observed 
properties of BLs, it inherits the dichotomization of 
BL Lac objects that was imposed by the two different discovery methods. 
Such a dichotomization can hinder our efforts to understand the physics
of BLs, particularly if the two methods of discovery force
us to observe the two extremes of a continuous distribution, thereby
 introducing  
 a  selection-induced bimodality. 

Strong indications that  a bimodality is introduced by the observational techniques
come from a recent study  by Laurent-Muehleisen (1997).
 The cross-correlation of the Rosat All Sky Survey (RASS) 
and the Green Bank 5 GHz radio survey 
revealed the existence of a previously highly undersampled 
population of BLs (120 objects) with properties intermediate 
between those of XBLs and RBLs. Specifically, the X-ray to radio flux ratio
of these sources and their location on a diagram of  radio to optical  
($\alpha_{ro}$)   versus optical 
to X-ray ($\alpha_{ox}$) spectral index bridge the gap between XBLs and RBLs.
Similar results (\cite{perlman98}) are emerging from  the deep X-ray radio blazar survey (DXRBS).
The question that naturally arises is whether there is an observed quantity
that varies in a continuous fashion, has a well defined physical significance,
and can be used as a classifying parameter.
We argue that such a quantity is the synchrotron peak frequency $\nu_{s}$.

\begin{figure}
\epsscale{.8}
\plotone{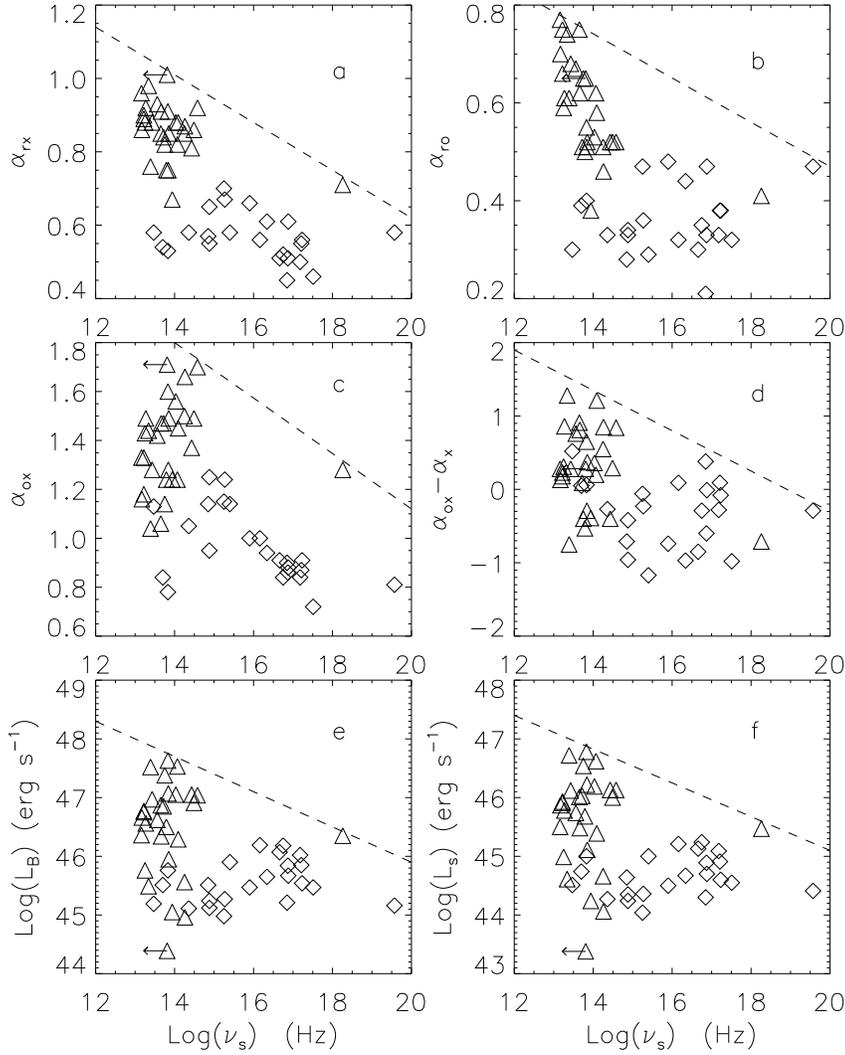}
\caption{The (a) radio to X-ray ($\alpha_{rx}$), (b) radio to optical 
($\alpha_{ro}$), (c)  optical to X-ray ($\alpha_{ox}$) broadband spectral indices;
(d) the X-ray concavity index $\alpha_{ox}-\alpha_{x}$; 
(e) the bolometric synchrotron apparent luminosity $L_{B}$; 
and (f) the synchrotron peak luminosity $L_{s}$,
as a function of the observed synchrotron peak frequency $\nu{_s}$ 
of the synchrotron  power spectrum  for the XBL (diamonds)
and  RBL (triangles) samples of Sambruna et al. (1996). 
The arrow indicate an RBL  with only an upper limit to   
 the peak frequency $\nu{_s}$. The dashed lines separate the populated from
 the unpopulated areas of the diagrams.} 

\label{npeak}
\end{figure}

Recently, Sambruna et al. (1996) examined the multifrequency spectral 
properties of three
complete samples of blazars: the \sl Einstein \rm Extended Medium Sensitivity 
Survey sample (EMSS) of XBLs (23 sources; \cite{morris91}), the $1 $ Jy 
sample of RBLs (29 sources; \cite{stickel91}), and a small complete sample of FSRQ from 
the S5 survey (8 sources; \cite{brunner94}). 
After they calculated the rest-frame 
fluxes and 
spectral indices for each object, they fit a parabola to the synchrotron 
power ($\nu L_{\nu}$) spectrum.  X-ray fluxes were excluded from the fitting 
in the cases  in which the X-ray spectral index $\alpha_{x}$ was 
flatter than the
broadband  optical  to X-ray spectral index $\alpha_{ox}$, a condition  that is
thought to  indicate  that   inverse Compton  radiation  
dominates  over the steep high-energy synchrotron tail of the spectrum. 
In Figure \ref{npeak}a--f we  plot
the spectral indices  $\alpha_{ox}$, $\alpha_{ro}$, $\alpha_{rx}$, the 
X-ray concavity index $\alpha_{ox}-\alpha_{x}$, and the synchrotron bolometric
 $L_{B}$ and peak  $L_{s}$ luminosities  as a function of the synchrotron peak 
frequency $\nu_{s}$ for the two BL  samples  of Sambruna et al. (1996). 
If we consider the two
samples collectively, we find that in all six diagrams there
is an upper envelope separating the populated area from a 
 well-defined zone of avoidance. The upper envelope is  such  that
for a given peak frequency $\nu_{s}$ there is a range of  
``avoided'' luminosities and spectral indices. As $\nu_{s}$ increases, the maximum observed
luminosity decreases, the steepest observed spectral index flattens, and the
maximum observed concavity index $\alpha_{ox}-\alpha_{x}$ decreases. 
On the other hand, the flattest spectral indices observed and the most negative concavity
indices do not seem to be very sensitive to the value  of  $\nu_{s}$. 

We propose a continuous classification system for BLs, based on the
synchrotron peak frequency  $\nu_{s}$ of the  synchrotron SED.  For example, 
a BL  peaking at $\log(\nu_{s})\approx 14$ will be classified as a BL14. 
The class of a BL Lac object then defines the maximum luminosity of the
 object, 
the maximum
steepness of its synchrotron spectrum over a range of  frequencies, and the
maximum concavity index. We note here  that (unlike the transient variability
mentioned before) long term, persistent changes
in $\nu_{s}$, should be reflected in the classification of the object.    

The synchrotron peak frequency $\nu_{s}$ of a BL  is  a very 
important parameter from a physical point of view as well.
 It is the frequency in 
which we receive the highest synchrotron power per logarithmic frequency 
interval. It is
also closely related to the peak of the inverse Compton emission in both the
SSC (\cite{bloom96}) and the 
EC models (\cite{sikora94}; \cite{dermer93}). This
frequency  is closely linked  to the the energetics and geometry 
of the synchrotron source 
and to the angle formed between the observer and the  plasma  
velocity, if  the source is moving relativistically 
(see \S \ref{results}). 

As Sambruna et al. (1996) point out, the determination of $\nu_{s}$
requires simultaneous multiwavelength flux 
measurements and can be particularly sensitive to a lack of flux measurements
 close to the 
peak of the distribution. The problem of multiwavelength coverage becomes more
difficult when $\nu_{s}$ happens to be either in the observationally 
inaccessible frequency range between sub--mm and infrared frequencies 
$\nu\approx 10^{12}-10^{14}$ Hz, or
in the UV due to galactic absorption. 
If  $\nu_{s}$ is in the  X-ray energy range, and there are no
flux measurements to constrain the parabola at frequencies higher than 
$\nu_{s}$, 
 the error in $\nu_{s}$ can be very large. This is the case 
for the two sources that have $\log\nu_{s} >  18$, the XBL 1433.5+6369,
and the RBL 2005-489 (see Fig. 4 of Sambruna et al. (1996)), which we 
decided not to consider further in our study.  
 In general, the error in the determination of $\nu_{s}$ 
should be smaller for objects with SEDs peaking  at IR/optical frequencies,
  where $\nu_{s}$  is more easily  constrained.

We note, however, that even though $\nu_{s}$ is 
difficult to measure exactly, 
an approximate estimate within  a decade of frequency is both feasible and 
adequate for the purpose of classification. Furthermore, the accuracy can be
improved if one decides to perform multiwavelength  observations with this
specific goal in mind, instead of using the available data from the literature.

Variability can also affect $\nu_{s}$. In the case of individual flares, 
initially 
the spectrum hardens (\cite{ulrich97}) 
and the peak synchrotron frequency increases. This change is temporary, and the 
source returns to its pre--flare  characteristics after the end of the flare.
The most extreme recorded
case of such a  shift occurred in Mkn 501, that increased its synchrotron
peak frequency by more than a factor of 100, changing form a BL17 to a
BL19 (\cite{pian98}). Usually, though, the shift in $\nu_{s}$, something 
neither  easily nor customarily calculated in multiwavelength
 studies, is
approximately  a factor of 10 in the case of the few well documented, 
large amplitude  flares. In the case of   Mkn 421  during
the May 1994 (\cite{macomb95}) and the April--May 1995 
(\cite{buckley96})  campaigns,  $\nu_{s}$ increased
by a factor $\approx10$--$100$, while during the PKS 2155-304 May 1994 
campaign,  $\nu_{s}$ increased by less than a factor of 10 (\cite{urry97}). 
Since these  excursions of  $\nu_{s}$  are usually of the same
magnitude as the uncertainty in measuring $\nu_{s}$, and also of 
a transient character, we do not consider them
a serious problem in assigning a specific $\nu_{s}$ to a BL, regardless of
variability.

Under our proposed  unified classification based on $\nu_{s}$,
 the questions that can direct us toward
a physical unification are:  (1) which are 
the physical and/or geometrical parameters that, 
given their range of values, 
create the range of the observed phenomenology?  (2) 
Can a model incorporate these
parameters and derive the properties of complete samples of BL Lac objects?
In the following section we propose  a simple unification scheme based on a 
self--similar jet  description.

\section{A  $ \bf \Theta$--$\bf \Lambda$ UNIFICATION SCHEME\label{unification}}

\subsection{A self--similarity hypothesis\label{basic}}

In order to  the extended region that BLs occupy
in the diagrams of Figure \ref{npeak},  we need to employ  a model with at 
least two free parameters. Any model with only one 
free parameter  can at best trace a curved trajectory in the observational
graphs when this free parameter is allowed to vary.  
Under the assumption that the jets of  BLs are relativistic,
a change in the angle between the line of sight and the jet axis will
induce a model-dependent change in the observed properties of the source.  
 It therefore seems necessary to postulate minimally 
 the existence of a physical free parameter that, together with
the angle, is capable of reproducing  the observed range of properties of 
 BLs.

The concept that relativistic jets can be described in a scale-invariant way
using  a self-similar description  is very 
appealing, since these physical configurations span a luminosity 
range greater than ten orders of magnitude,  from the  jets in stellar
systems in our galaxy to the
most extreme FSRQs . 
Maraschi \& Rovetti (1994), after studying samples of BLs, FSRQ, 
and FRI
and FRII radio galaxies, suggested that all these sources operate in a self 
similar way, with the jet luminosity and angle between the line of sight 
and the jet 
axis accounting  for the different observed properties. 
Gear (1993) studied the infrared colors 
of two complete samples of RBLs and XBLs, and hinted 
 toward a BL  unification based on the  luminosity and the 
angle between the line of sight and the jet axis. 
 
We propose  that the second, physical  
unification parameter is the  
electron 
kinetic luminosity of the jet $\Lambda_{kin}$. At first view,  one would be 
tempted to assume that  $\Lambda_{kin}\propto r_{\star}$, since the
Eddington luminosity scales with the black hole mass,  which is proportional
to the gravitational radius of the black hole.
However, tying  the luminosity of the system to the black hole mass 
cannot address
the evolutionary characteristics of BLs. The black hole mass  
slowly increases, due to accretion of fresh material,
and this  translates into a  slow increase of  the luminosity,
  which essentially translates  to a weak 
negative  cosmological evolution, i.e.,  each source becomes more powerful 
as cosmic time increases. We know that this
is  the  case neither  for RBLs (see \S\S \ref{evolution}), nor  for all the other
 radio--loud AGN families (\cite{urry95}),
which we presume are described by the same paradigm --- a massive black hole,
an accretion disc, and a relativistic jet.

An alternative description has been  proposed by 
Rees et al. (1982). Under this  scheme,  the electromagnetic power $L_{EM}$
 extracted 
from of a rapidly
rotating black hole is proportional to $a^2$,  where $a<r_{g}$ is a length 
that measures the specific angular momentum
$ac$ of the black hole, and $r_{g}$ is the gravitational radius of the 
black hole.
Assuming now  that $L_{EM}$ and $a$ are related linearly  to 
  $\Lambda_{kin}$ and $r_{\star}$ correspondingly, 
 we propose that  $\Lambda_{kin}$ scales as
$r_{\star}^2$:
\begin{equation}
\Lambda_{kin}\propto r_{\star}^2 .
\end{equation}
Under this scheme, the main cause for evolution is
 the spin down of the
black hole, manifested as a  decrease of 
$a$ due to the continuous energy extraction. This way the 
power of the source
 decreases with cosmic time, giving rise to the observed positive 
cosmological evolution, in the sense that
 sources were more powerful in the past (the apparent 
negative evolution of XBLs is 
addressed in \S\S \ref{evolution}).

We additionally assume that  the intensive  physical variables  that describe 
the relativistic
jet in BLs, such as  the magnetic field, 
the comoving plasma density, and the high energy cutoff of the injected 
electron energy distribution, all vary over  a small intrinsic 
range of values. The 
combination of these two assumptions suggests  a unification scheme in which 
the observed properties of a BL 
depend mainly on the electron kinetic luminosity $\Lambda_{kin} $ 
of the jet and the angle between the line of sight and the jet axis
$\Theta$.
A similar scaling relation  has been discovered observationally 
 between the size
 $R_{BELR}$
of the broad emission line region (BELR)  and the central (accretion disk) 
ionizing luminosity $L_{acc}$
of Seyfert galaxies and quasars  (\cite{kaspi96}) using reverberation
 mapping:
\begin{equation}
L_{acc}\propto R_{BELR}^2 .
\end{equation}
 This scaling law implies
an ionization parameter $U$ (ratio of ionizing photon density to electron 
density) that has a small intrinsic range of values ($U\sim 0.1-1$)
for objects that extend in luminosity over almost four orders of magnitude
(\cite{wandel97}). Observational arguments, suggestive  of a self--similar 
description of radio-loud AGN, have been presented by 
Rawlings and Saunders (1991), who find that the kinetic luminosity of the 
kpc-scale jet is proportional to the narrow emission line luminosity.
 This relation holds
for objects with or without broad emission lines and extends for more than 
four orders of magnitude in luminosity. 
In a similar study, Celotti, Padovani, 
\& Ghisellini (1997) find that  $ L_{BELR} $ 
and the pc-scale
kinetic luminosity of the jet for a sample of radio loud AGN are of the same
order of magnitude. 
As recent
studies of radio loud AGN suggest (e.g. \cite{koratkar98}), 
the BELR is  probably ionized by thermal
radiation from the central engine, rather than by the jet radiation.

We adopt as a working hypothesis a picture in which the 
ionizing luminosity of the accretion disk  
$L_{acc}$, the jet electron kinetic luminosity $\Lambda_{kin}$, and the BELR  
luminosity  $L_{BELR}$  are closely related,  scaling as
\begin{equation} 
L_{acc}\propto L_{BELR} \propto \Lambda_{kin} \propto r^{2}_{\star}, 
\label{eq:scale} 
\end{equation} 
where $r_{\star}$ is a dimension  that characterizes the jet. 
We stress here that
this 
scaling assumption  does not refer to a specific jet model, but is a
rather general, model independent  hypothesis that can be applied to  
specific jet models.

\subsection   {Application to the accelerating jet model \label{apply}}

\begin{figure}
\epsscale{0.8}
\plotone{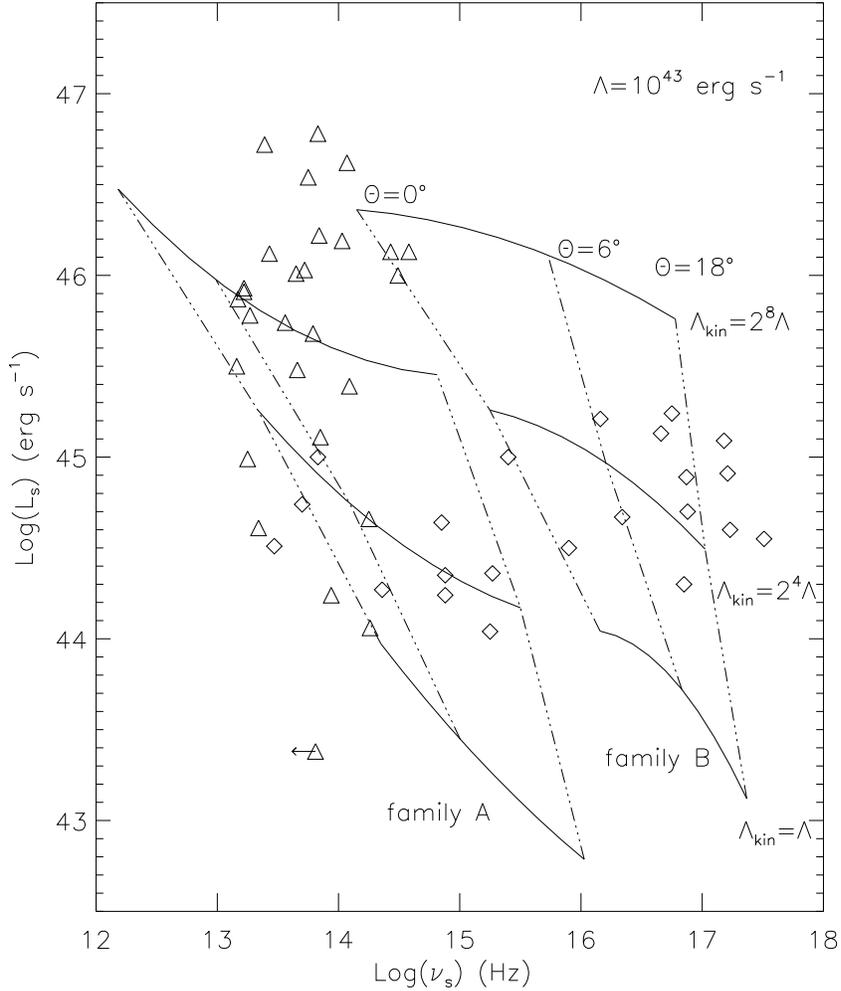}
\caption{The synchrotron apparent peak luminosity $L_{s}$ 
as a function of the observed synchrotron peak frequency $\nu{_s}$ 
of the SED for the XBL (diamonds)
and  RBL (triangles) samples of Sambruna et al. (1996). The two superposed
grids are the results of the accelerating jet model coupled with 
the scaling  assumption of equation (\ref{eq:scale}) 
for two different physical descriptions
of the jet, as given  in Table \ref{tbl3}.
 The solid curves represent constant electron kinetic luminosity 
$\Lambda_{kin}$, while the dash--dot curves  correspond to
 constant angle $\Theta$ 
between the jet axis and the line of sight.}
\label{uni}
\end{figure}

We apply  this self--similar scheme to the accelerating jet model.
The question we  address here is whether we can populate the  $\nu{_s}$--$L_{s}$ 
diagram of Figure 
\ref{npeak}f, under the assumptions that  the jet  electron kinetic  
luminosity scales with the square of the  jet size, and that the physical
parameters which describe the jet have a small intrinsic range of values.
We perform calculations using   two somewhat different families  of values for the 
physical parameters,
listed  in Table \ref{tbl3}. For each family we run 
the simulation for a range 
of angles and luminosities, scaling
the square of  $r_{\star}$ and  $z_{\star}$ 
with the electron kinetic luminosity $\Lambda_{kin}$. The results of this set of 
simulations can be seen in  Figure \ref{uni}. The solid curves 
represent calculated values of $L_{s}$  for a constant electron kinetic luminosity and size
as the angle $\Theta$ varies between $0^{\circ}$ and $18^{\circ}$. The 
dash--dot curves 
correspond to  constant angle as the electron kinetic luminosity
$\Lambda_{kin}$ increases  from
$\Lambda$ to  $256  \Lambda$, where $\Lambda=10^{43}$ erg s$^{-1}$. 
This increase in luminosity is accompanied by an increase in the jet size:
$r_{\star}$ and $z_{\star}$ increase by a factor of 16 , starting at 
 $r_{\star}= 2.5 \times 10^{14}$ cm  and  $z_{\star}= 5.0 \times 10^{14}$ cm.

The qualitative behavior of both model families is the same. For a given 
electron kinetic luminosity, as the   angle $\Theta $ increases, the peak frequency
$\nu_{s}$ increases and the peak apparent luminosity  $ L_{s} $ decreases. 
For a constant angle, as $\Lambda_{kin}$  decreases, there is a 
drastic decrease in $L_{s} $  and a milder
increase  
in  $\nu_{s}$. 
The two families of models generated by these somewhat different 
physical descriptions cover a significant
portion of the observed range of values. 
This behavior is not specific to the above
choice for the physical parameters: a different set of jet parameters 
would generate  a pattern similar to those  obtained by the models of Fig.
\ref{uni}/Table \ref{tbl3}.
Modest  changes in the physical description merely move the grid around the 
observed parameter space. A small range
of values for the physical description of the jet, coupled with  the
 $\Theta$--$\Lambda$  scheme, can populate the $\nu_{s}$--$L_{s}$ 
diagram for BLs. Choosing the appropriate set of jet parameters, the  
$\Theta$--$\Lambda$  scheme can produce synthetic spectra of BL18--BL19
sources. We choose not to do so, since the determination of $\nu_{s}$
for the two sources with $\log\nu_{s} \geq 18 $ in Sambruna et al. (1996)
is problematic. 

The conclusion reached previously by Sambruna et al.
(1996) and Georganopoulos and Marscher (1996) that the range of $\nu_{s}$ 
cannot
be explained merely by jet orientation is supported by this study. However,
a  combination  of jet orientation
and a relatively small range in the physical parameters that describe the jet
can explain the range of $\nu_{s}$ under the $\Theta$--$\Lambda$  scheme.
A quantitative comparison, including specific
assumptions about the range of the physical parameters and extensive 
comparison of  synthetic samples to existing samples, 
does not seem worthwhile until the parameters can be more strictly 
constrained, for example through analyses of multiwavelength variability.

One  might naively expect that, for a given angle, 
since  the jet electron kinetic luminosity 
and  size scale in a  self--similar fashion, the SED would follow this 
self--similar  behavior by scaling up or down, following the electron kinetic 
luminosity and  
keeping its shape  and peak frequency the same.  
An important feature  emerging from our  simulations is that
for a constant angle,  as the electron kinetic luminosity and the size of the jet
increase, the peak  frequency decreases. This means that, although the jet
scales in a self--similar manner,  this self--similarity is not transfered
completely to the 
SED, since the peak frequency  changes. 
The reason  for this break-down is that the effects of synchrotron losses 
(cf. eq. \ref{eq:dens} and the discussion of \S\S \ref{parspace}) 
on $\nu_{s}$ do not follow scaling laws that are qualitatively similar to
those that govern the behavior of $L_{\nu}$ at $\nu \leq \nu_{s}$.
For example, if the values of $r_{\star}$ and  $z_{\star}$ were doubled while
holding $B_{\star}$ constant, $\nu_{s}$ would decrease because the electrons
would suffer more severe synchrotron losses as they traverse the characteristic
distance $z_{\star}$ before encountering the region of the jet where B 
decreases.

\subsection{The negative  evolution of   XBLs \label{evolution}}
The evolution of all the  AGN families is positive, in the sense that these
objects were either more  common or more luminous (or both) in the past. 
A unique and striking exception 
is the negative evolution of XBLs:  Morris et al. (1991) showed that  
the evolution of the EMSS XBL sample is negative. It seems that these
objects  
were either less common or less luminous in the past.  
This has been verified recently by Perlman et al. (1996) for the ROSAT
PSPC 
sample of XBLs, with $\langle V_{e}/V_{a}\rangle=0.331 \pm 0.060$, 
where $V_{e}$ is
the  
volume of a sphere with radius equal to the distance to  each object and $V_{a}$ is the available volume
within  
which each object could have been detected in the EMSS. This result  
is in  contrast to the value $\langle V_{e}/V_{a}\rangle=0.60\pm0.05$ 
obtained  by Stickel et al. 
(1991) for the 1 Jy RBL sample. 

This negative evolution of XBLs can be explained naturally  in the
framework 
of the $\Theta$--$\Lambda$ scheme, under the assumption commonly  invoked  for
AGN  that
the 
electron kinetic luminosity
$\Lambda_{kin}$ of the radio--loud AGN jets  has a positive evolution, 
viz. jets (on average) in AGN   had a 
higher  $\Lambda_{kin}$ in the past.  As we showed in \S\S \ref{apply} 
for a given set of 
physical parameters and angle $\Theta$, as $\Lambda_{kin}$ decreases, 
the peak  
synchrotron apparent luminosity $L_{s}$ decreases and  
the peak synchrotron frequency $\nu_{s}$ increases,  essentially shifting  
the source  
from the region of the  RBL-type objects  
toward that of the  XBL-type objects. If we start with a sample 
of sources characterized by a range of $\Lambda_{kin}$ and $\Theta$ and  
let these 
sources evolve to lower values of electron kinetic  luminosity $\Lambda_{kin} $,
 then each source will gradually (with cosmological time) shift 
its characteristics to become more
XBL-like.  This means that  the fraction of sources that would be  
classified 
as RBLs will gradually decrease with time/redshift,  resulting in 
 $\langle V_{e}/V_{a}\rangle  \gtrsim 0.5$ in a radio-flux 
limited sample, while the fraction of sources that would be  
classified 
as XBLs will gradually decrease and yield  
$\langle V_{e}/V_{a}\rangle \lesssim 0.5$ in an X-ray flux-limited sample.  

Our model predicts  that the  value of  
$\langle V_{e}/V_{a}\rangle$ of an intermediate BL sample, such as that  of 
 Laurent-Muehleisen (1997), will be confined between the  corresponding values
of $\langle V_{e}/V_{a}\rangle $ for the XBL and RBL samples mentioned above,
exhibiting  
almost no   evolution ($\langle V_{e}/V_{a}\rangle \approx 0.5$), since the
population of intermediate BLs on one hand will gain members from the RBLs,
and will lose members to the XBLs as cosmic time increases.  
In fact, a very recent
report (\cite{bade98})  agrees with this prediction. Bade et al. (1998)
have studied a new sample of XBLs selected from  RASS. They subdivide their
sample into two halves according to the X-ray to optical flux ratio 
and find that
the extremely X-ray dominated subgroup shows negative evolution, while the
subgroup with intermediate SEDs is compatible with no evolution at all.
They also find, however, that the extremely X-ray dominated subgroup has 
higher X-ray
luminosity and redshift than the intermediate subgroup, an effect  that
does not seem to agree with  our model.

\subsection{Emission line strength under the $ \bf \Theta$--$\bf
\Lambda$ scheme
 \label{lines}}
The discussion in \S\S \ref{basic} implicitly treats BLs as AGNs with a BELR
and a central isotropic source of thermal ionizing radiation, essentially 
reducing 
 the differences between BLs and FSRQ. The 
original definition of a BL Lac object 
includes the criterion of absent or weak
(equivalent width $W_{\lambda} \lesssim 5$ \AA) { }
 emission lines (\cite{stickel91}). This
has led most workers in the field to assume that either the gaseous 
environment
near the central engine of BLs is very poor, or there is no 
significant source of ionizing radiation, or both. Additionally,
  the conviction
of a bimodality separating BLs and FSRQ in terms of their emission
line properties had been established.
 The situation, though, is far from clear and recent results  
point toward a continuous transition from BLs to 
FSRQs in terms of their emission line characteristics. Even in the complete
sample of Stickel et al. (1991), the authors identified 6 out of 34 RBLs that
had emission lines violating the equivalent width limit of $5 $ \AA. 
It is also true  that the equivalent width is  variable and 
there are objects that  change classification depending on the time of 
observation (\cite{antonucci93}; \cite{ulrich97}).

 The BL PKS 0521--365 
(\cite{scarpa95}), and the prototype object of
this class, BL Lacertae (\cite{vermeulen95}),  have shown emission lines 
with equivalent widths $W_{\lambda} \gtrsim 5 $\AA { }. More surprisingly,
the X-ray selected BL  Mkn 501, which has recently been detected  
at TeV energies (\cite{quinn96}), has shown emission lines  in the past
(\cite{moles87}).
Corbett et al. (1996) argue in  the case of BL Lacertae
 that the most plausible  way to power 
the emission lines is through thermal radiation from an accretion disk. 
The required thermal radiation power does not alter significantly the total 
observed flux and the object does not show the  ``big blue bump''
component that many radio loud quasars exhibit.

In a recent study of emission-line luminosities for a sample of 
HPQs and RBLs, Scarpa \& Falomo (1997) find 
that the line luminosity ranges of HPQs and BLs largely overlap, 
although 
RBLs have smaller equivalent widths. Specifically, in Figure 10 of 
their paper they plot
the line luminosity in Mg II versus the continuum luminosity for the sources 
of their sample, as well as the line that corresponds to the equivalent 
width limit $W_{\lambda}=5$  \AA. Their results clearly demonstrate that
 for any given
continuum luminosity the corresponding emission line luminosity does not
show any gap separating RBLs and HPQs. If we were to increase the arbitrary
 $5 $ \AA { }
 limit then we would 
 accept as BLs sources that previously have been classified as HPQs.
Similarly, lowering the equivalent width limit would  shift objects
from the BL  to the HPQ family. 
It therefore seems  plausible that the
gaseous environment of BLs, although poorer than that of the HPQs, may not 
 be qualitatively different.

Additional arguments for the presence of ionized  gas
close to the central engines of BLs comes from studies of 
X-ray and EUV absorption features in XBLs. Madejski et al. (1991) 
found in all objects from   a sample
of five XBLs  an absorption feature at an energy of
$\sim 650$ eV. K{\"o}nigl et al. (1995) studied the X-ray bright BL
PKS 2155--304 with the Extreme Ultraviolet Explorer and 
found an absorption feature between  $\sim$ 75 and $\sim 85 $ \AA. They modeled
this feature as absorption resulting from gas clouds at a distance from the
central engine similar to the typical BELR size, moving with a speed  
$\sim 0.1$ c.  Similar results have been reported recently from EUV 
observations of the XBL Mkn 421 (\cite{kartje97}),
 and X-ray observations of the XBL H1426+428 (\cite{sambruna97}).

\begin{figure} 
\epsscale{0.8} 
\plotone{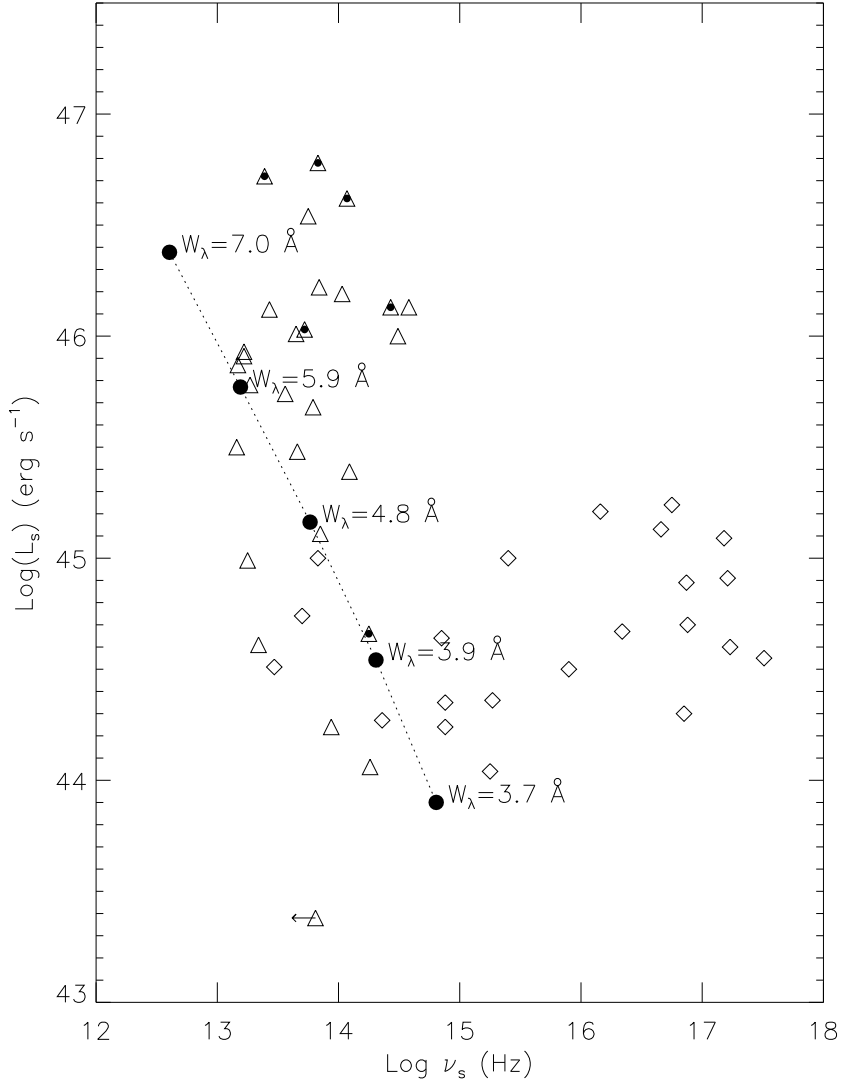} 
\caption{The synchrotron apparent peak luminosity $L_{s}$ 
as a function of the observed synchrotron peak frequency $\nu{_s}$ for 
 the XBL (diamonds) and RBL (triangles)
samples of Sambruna et al. (1996). Small filled circles mark
 RBLs with emission lines that
violate the $ 5 $\AA { } equivalent--width criterion. The bigger filled circles, 
connected with a dotted line, represent the model points of Figure \ref{line}.
The most luminous model source is assigned an arbitrary equivalent
width $W_{\lambda}=7$\AA { }, and the variation of the equivalent width 
$W_{\lambda}$ is shown as $\Lambda_{kin}$ and $L_{BELR}$ decrease.} 
\label{line2} 
\end{figure}

In Figure \ref{line2} we plot the synchrotron peak apparent luminosity $L_{s}$ 
as a function of the synchrotron peak frequency $\nu{_s}$ for 
 the XBL (diamonds) and RBL (triangles)
samples of Sambruna et al. (1996). The RBLs with emission lines that
violate the $ 5 $\AA { }  equivalent width criterion are marked with  small
filled circles. Note that these sources are preferentially low $\nu_{s}$,
high $L_{s}$  objects.
We use the scaling relation  
(\ref{eq:scale}), 
coupled with the accelerating jet model  
to show   that the lack of emission lines  in high $\nu_{s}$, 
low $L_{s}$  (BL15--BL17) BLs  
and the existence of emission lines preferentially 
in  low $\nu_{s}$,  
high $L_{s}$  (BL13--BL14) BLs can be reconciled with the
existence of 
 a scaled-down accretion disk illuminating a BELR. 

\begin{figure} 
\epsscale{0.8} 
\plotone{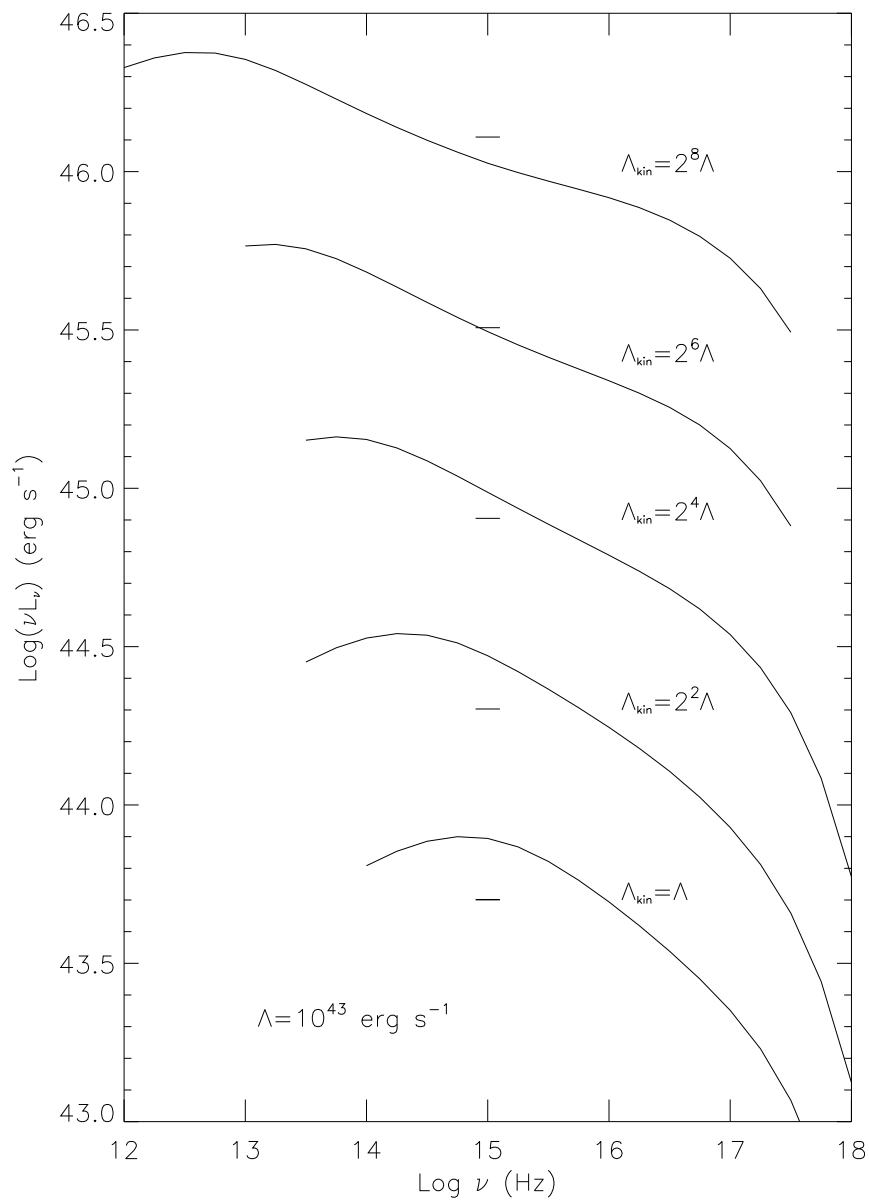} 
\caption{The SED for a range of jet electron kinetic luminosities. The BELR
luminosity, shown by the short horizontal line is proportional to the jet 
electron kinetic luminosity (see \S\S \ref{lines}).
The parameters that describe the jet are the same 
as the parameters describing the family A jets in Table \ref{tbl3}, 
with $\Theta=3^o$.} 
\label{line} 
\end{figure}

 In Figure \ref{line} we plot the SED for a range of
$\Lambda_{kin}$. We assume that the BELR luminosity is proportional  to
the jet electron kinetic luminosity, and for the purpose of demonstration
we consider a broad line with a full width at half maximum (FWHM) of 3000 km/s and   with power
 $L_{BELR}=0.3  \Lambda_{kin}$. The BELR
luminosity  is shown by the short horizontal 
line centered at $\nu=10^{15}$ Hz. As $\Lambda_{kin}$ and 
 $L_{BELR}$ decrease, the synchrotron peak frequency $\nu_{s}$
increases. Although the decrease of $\Lambda_{kin}$ is accompanied by
a decrease in the emitted power $\nu L_{\nu}$  at every frequency, 
the shift  of $\nu_{s}$ toward higher frequencies 
 results in the gradual dominance of the 
synchrotron emission over the line radiation.
In Figure  \ref{line2} we plot as filled circles 
the loci of the model for the range of
luminosities used in Figure \ref{line}. 
 The qualitative behavior of the model is similar 
to the trends
exhibited by the data: as  the model points shift  to higher values
of $\nu_{s}$ and lower values of $L_{s}$, the equivalent width
$W_{\lambda}$ decreases, crossing over to values smaller than $ 5 $\AA.

Thus, although the ratio of jet electron kinetic luminosity to BELR luminosity
remains the same, the equivalent width of the emission lines
decreases as the  electron kinetic luminosity of the source decreases.
This is again a manifestation of the breakdown in self-similarity 
described  in \S\S  \ref{apply}.

 \section{SUMMARY AND DISCUSSION \label{discussion}}

 The main points of this work are as 
follows.

1. We verify that an increase in the angle  $\Theta$ between the line 
  of sight
 and the jet axis  shifts the observed characteristics of the source
 from the RBL to the XBL class. As  has been noted, however
 (\cite{sambruna96}; \cite{markos96}), this shift is
 not enough to explain the range of observed physical parameters in 
 complete BL  samples. 

2. We propose a new continuous classification scheme for BLs, based
 on 
  the synchrotron peak frequency $\nu_{s}$ of the SED of a source. 
 This scheme avoids
 the observationally induced bimodality of the currently 
 used XBL--RBL, or HBL--LBL schemes, and 
 naturally accommodates the newly discovered (\cite{laurent97}; 
\cite{laurent98}; \cite{perlman98})
 intermediate BLs.
 As  can be seen in the results of Sambruna et al. (1996), given the
 class of a BL, we can infer  upper limits to the synchrotron 
 luminosity of the object, the steepness of the spectrum at different
 frequencies, and the maximum concavity index $\alpha_{ox}-\alpha_{x}$.

3. Observations suggest  that BLs may be characterized
 by a scaled-down accretion disk and BELR, similar to those found in FSRQs.
Based on recent  observational results, we
propose a self-similar blazar  description, according to which the
 accretion disk luminosity, the BLR luminosity, and the jet kinetic luminosity
 scale with the square of the scale size of the jet
[expression (\ref{eq:scale})].
 This description leaves
 invariant all the primary  physical parameters 
such as the pressure and particle and energy density at the base of the jet. 
 
We propose the $\Theta$--$\Lambda$ scheme unification scheme:
the electron kinetic luminosity, together with  the orientation of the jet,
determines the observed characteristics of a source. A small range of the
physical parameters that describe the jet, together with the 
 $\Theta$--$\Lambda$ scheme, can reproduce the range of observed BL synchrotron
peak frequencies $\nu_{s}$, which  cannot be reproduced solely by
orientation effects  (\cite{sambruna96}; \cite{markos96}).

 4. The negative evolution of the XBLs can be explained naturally 
  in the context of this self-similarity based scenario.
 If we start with a random sample
 of sources, as cosmological time passes the electron kinetic luminosity decreases, 
 the peak frequency $\nu_{s}$ for each source increases, and  the number
 of sources that would be classified as XBLs increases. We predict that
 samples of  intermediate sources will have a $\langle V_{e}/V_{a}\rangle$
 value intermediate
 between those of XBL and RBL samples, as new results have indicated.

 5. We show that the existence of emission lines preferably
 in low $\nu_{s}$ (BL13--BL14), high $L_{s}$  sources, and the lack of 
 lines in higher $\nu_{s}$, lower $L_{s}$  objects, is reproduced under
 scaling relation (\ref{eq:scale}). As both  $\Lambda_{kin}$  and 
  $L_{BELR}$ decrease, $\nu{_s}$
 shifts toward the frequency of the emission lines and  the non--thermal
 continuum  gradually dominates over  the emission line luminosity. 

 Our discussion of  the $\Theta$--$\Lambda$ scheme has been confined to the
 BLs, although there are many indications that a common
 underlying physical mechanism could describe all blazars. 
 Since the  luminosity output of FSRQs is 
 dominated in many cases by the $\gamma$-ray flux and  our simulation does 
 not calculate
 the inverse Compton energy losses, we cannot at present  apply our 
model to  FSRQs. 
  The extended radio luminosity distribution of FSRQs is similar to
  that of  FR II radio galaxies (\cite{maraschi94}), and
 the jet magnetic field tends to be aligned with the jet axis (\cite{cawthorne93}).
 BLs, on the other hand,  have an extended radio luminosity distribution 
 similar to  that of  
  FR I radio galaxies (\cite{perlman93}),
 and the jet magnetic field tends to be  perpendicular to the jet axis 
 (\cite{gabuzda92}). 

Although it may  seem as if  these differences are enough to separate BLs 
and FSRQs into two distinct families,
 there are many observational arguments that
 support the possibility of a continuous distribution of sources that
belong to a single, continuous family. We have already reviewed observations
  suggesting that a  separation of blazars
 on the 
basis of the emission line properties may not be justified. 
The extended radio power distribution
for BLs and FSRQs overlap significantly, and the same holds for  
their parent populations, the FR I and FR II radio 
galaxies (\cite{maraschi94}; \cite{padovani92}).
A population of FR II radio galaxies with low-excitation narrow-line 
spectra,
 low radio power, and
more relaxed radio morphologies  (\cite{laing94}) is indicative of a 
connection between FR I and FR II sources. Recent VLBI studies 
(\cite{cawthorne96};
Kollgaard et al. 1996; \cite{gabuzda94};
\cite{leppanen95}; \cite{kemball96}) indicate that the 
magnetic field orientation does not always follow the previously observed
bimodality, and FSRQs with perpendicular and BLs with parallel
magnetic field are not rare. Additionally, recent results (\cite{lister98})
from higher frequency (43 GHz) VLBI observations
indicate that the  magnetic field orientations of both BLs
 and FSRQs in the radio core region are similar.

\begin{figure}
\epsscale{0.7}
\plotone{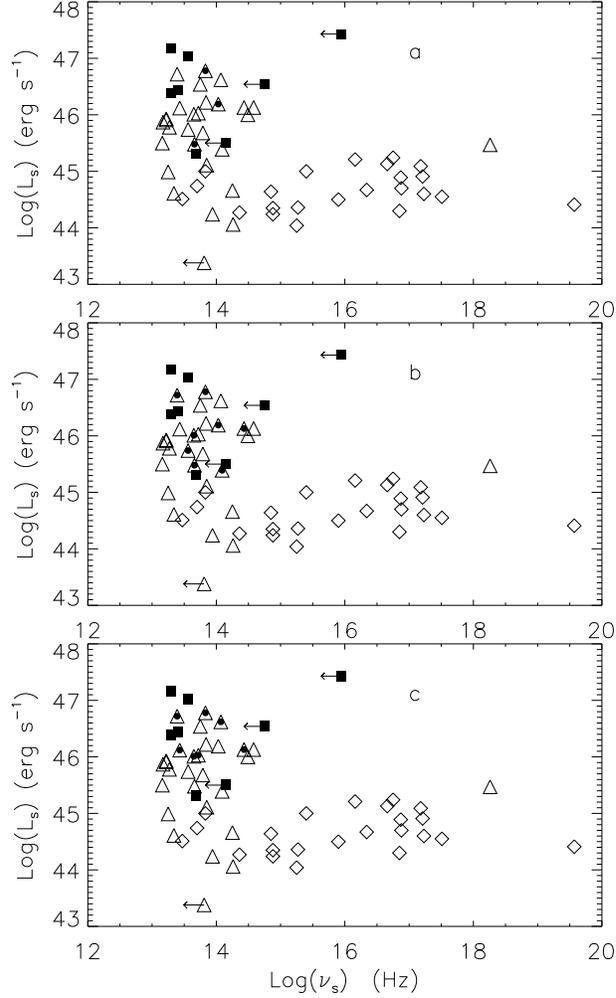}
\caption{Peak synchrotron apparent luminosity vs. observed peak frequency for
the two BL  samples (see Fig. \ref{line2}) and  the  small complete sample of FSRQ 
(filled squares) from the S5 survey (Brunner et al. 1994), used by
 Sambruna et al. (1996). Small filled circle mark: 
(a) RBLs with  magnetic field parallel to the jet
 (0735+178, 1308+326, and 
2007+777 from Gabuzda et al. 1994);
(b) RBLs with FR II--like extended radio 
characteristics (0235+164, 1308+326, 1538+149, and  1823+568 
from Murphy et al. 1993,
0954+658, 1749+701, and 2007+777 from  Kollgaard et al. 1992, and 1807+698	from
Wrobel et al. (1990);	
(c) RBLs  with emission lines that
violate the $5 $\AA { } equivalent width criterion (0235+164, 0537-441, 
0851+202, 
1308+326, and 1749+096 from Stickel et al. (1991), 
and  BL Lacertae from Vermeulen
et al. 1995).}

\label{bl}
\end{figure}

There are several
BLs in the samples that Sambruna et al. (1996) consider which 
have  extended radio characteristics that  correspond to  FR II 
radio galaxies (\cite{kollgaard92}; \cite{murphy93}; \cite{wrobel90}),
or have longitudinal magnetic fields (\cite{gabuzda94}).
 As  can be seen in Figure 
\ref{bl}, 
all these BLs are low $\nu_{s}$
(BL13--BL14), high $L_{s}$  objects
 that  occupy a region of the parameter space in the  
$\nu_{s}$--$L_{s}$ diagram  similar to that occupied by the 
FSRQ sample. 
It  therefore seems justifiable to consider
that the properties of BLs and FSRQs
 do not correspond  to two distinct physical 
structures, but, rather, form  a continuum.

Since the jets in these sources 
are relativistic, the angle that the jet forms with the line of sight must
be one of the parameters that affects their observed properties. 
The range of observed luminosities for objects with similar synchrotron
peak frequencies $\nu_{s}$ has led us to propose that a
second crucial parameter is  the
electron kinetic luminosity of the jet. 
Our simulations show that this formulation indeed holds promise
for a global understanding of blazars.
For further progress,  detailed, time--dependent  simulations 
 that include in an accurate  fashion 
the relevant 
physical processes, such as inverse Compton scattering of both jet
synchrotron photons and external photons, need to be developed and 
compared to observations of individual sources and to the statistical
properties of complete samples.
\acknowledgments This research was supported in part by the NASA Astrophysical Theory 
Program grant NAG5-3839.

\appendix
\section{GAS DYNAMICS \label{gas}}

We describe here the analytical one--dimensional steady flow described  by 
Blandford \& Rees (1974). The plasma produced in the vicinity of the central
engine is ultrarelativistic in the fluid proper frame  (mean total to 
 rest mass ratio per particle  $\gamma \gg 1)$, and remains in the 
ultrarelativistic regime as it flows downstream. The flow is 
considered adiabatic and quasi--spherically symmetric, i.e., that the jet 
is confined to a cone.
 The pressure profile is parameterized
by a power law,
\begin{equation}
p(z)=p_0\left(\frac{z}{z_0}\right)^{-a}
\end{equation} 
where the $z$ axis is the symmetry axis of the system, $z_0$ is the distance
of the flow stagnation point (bulk motion Lorentz factor $\Gamma=1$) from a
fiducial point where  the pressure mathematically becomes infinite, 
$p_0$ is the pressure at  $z_0$, and $a$ is the exponent
that describes how fast the pressure drops ($a \geq 0$) as we move away 
from the central engine along the  $z$ axis. Note that  $z=0$ at the fiducial 
point and not at the central engine. 
The jet is characterized by a  constant power,
\begin{equation}
\Lambda_{t}=\Gamma^2\beta w c A,
\label{eq:lambda}
\end{equation}
where $\beta$ is the jet velocity in units of $c$, $w$ 
is the enthalpy density in the fluid frame and
$A$ is the cross-section of the jet at a given $z$. For a relativistic gas
$w=(4/3)e$, where $e$ is the comoving energy density of the plasma.

 One way to ensure  that  the flow is adiabatic is 
to assume that a significant portion of the energy is 
carried by relativistic protons that have essentially no radiative losses.
Alternatively, we confine our study to EED 
with $s>2$, which corresponds to  most of the energy being near 
 the low energy cutoff of the 
EED. Since the low energy electrons have small radiative  losses, such a  flow 
is 
approximately adiabatic.
We use the first description:
\begin{equation}
e=e_{el}+e_{p},\; \;e_{el}=k\,e,\;e_{p}=(1-k)e,\; 0<k<1, 
\end{equation} 
where
\begin{equation}
e_{el}=\int_{\gamma_{min}}^{\gamma_{max}}\gamma N_{\star}\gamma^{-s}d\gamma=
\frac{N_{\star}}{2-s}(\gamma_{max}^{2-s}-\gamma_{min}^{2-s})
\label{eq:e}
\end{equation} 
is the electron energy density and $e_{p}$ is the proton energy density
in the comoving frame.
Assuming equipartition between the electron and proton energy density,  we set
$k=0.5$ (the actual requirement is $k \lesssim 0.5$). This ensures that half of the energy is in protons that do not
radiate, and therefore our adiabatic approximation is reasonable.
Therefore, the electron kinetic luminosity  $\Lambda_{kin}$ is
\begin{equation}
 \Lambda_{kin}=k\Lambda_{t}=\Lambda_{t}/2,
\end{equation}

The cross-section
of the jet is minimized at the sonic point ($\Gamma_{\star}=\sqrt{3/2}$),
which is characterized by a pressure 
\begin{equation}
p_{\star}=\frac{4}{9}P_0,
\end{equation}
a cross-sectional radius $r_{\star}$
\begin{equation}
r_{\star}=2^{1/4}\left(\frac{3}{2}\right)^{3/4} \left(\frac{\Lambda_{kin}}{4\pi c p_0}
\right)^{1/2},
\end{equation}
and a distance $z_{\star}$ from the fiducial point 
\begin{equation}
z_{\star}=z_0\left(\frac{9}{4}\right)^{1/a}.
\end{equation}

The cross-sectional radius of the jet as a function of $z$ is given by
\begin{equation}
r(z)=r_{\star}\left(\frac{z}{z_{\star}}\right)^
{\frac{3\epsilon}{2}}\left[3\left(\frac{z}{z_{\star}}\right)^{2\epsilon}-
2\right] ^{-\frac{1}{4}}
\label{eq:radius}
\end{equation}
and the bulk  Lorentz
 factor of the flow $\Gamma$ is given by
\begin{equation}
\Gamma(z)=\Gamma_{\star}\left(\frac{z}{z_{\star}}\right)^{\epsilon}.
\label{eq:Gamma}
\end{equation} 
The exponent $\epsilon$ determines how fast the jet opens
and accelerates and is determined solely by the steepness of the
external pressure gradient, $\epsilon=a/4$.
For $z\gg z_{\star}$, $r\propto z^{\epsilon}$. 

\section{THOMSON LOSSES \label{thomson}}
Our treatment of electron losses does not consider the Thomson losses
due to the radiation from an accretion disk. If $z_{1}$ is the distance
of the fiducial point used in the flow description, the distance of
the base of the jet from the central engine is $z_{base}=z_{1}+z_{\star}$.
The synchrotron losses will be greater than the Thomson losses 
 if the accretion disk photon energy density in
the comoving frame at the base of the jet is greater than the comoving 
magnetic field energy density.

As given in  Dermer \& Schlickeiser (1994), the comoving photon energy 
density $u_{acc}$ at distance $z$ from the central engine due to a 
point--like 
accretion disk is 
\begin{equation}
u_{acc}=\frac{L_{acc}}{4\pi z^2c\Gamma_{\star}^2(1+\beta_{\star})^2},
\end{equation}
where $L_{acc}$ is the accretion disk luminosity, and 
$ \Gamma_{\star},\; \beta_{\star}$ describe the jet flow at the base of
the jet. Requiring $u_{acc} < B_{\star}^2/(8\pi)$ we obtain
\begin{equation}
z_{base} >z_{min}=(2/3)^{1/2}10^{17} L_{44}^{1/2}B_{\star}^{-1} \Gamma_{\star} ^{-1}(1+\beta_{\star})^{-1}\mbox{ cm},
\label{eq:zbase}
\end{equation}
where $L_{44}$ is the accretion disk luminosity in units of 
$10^{44}$  erg s$^{-1}$.
In this work we assume that the fiducial point is adequately far from the 
central engine so that relation \ref{eq:zbase} is satisfied.

\appendix

\clearpage

\begin{deluxetable}{cc}
\footnotesize
\tablecaption{Model parameters. \label{tbl1}}
\tablewidth{0pt}
\tablehead{
\colhead{Parameter} & \colhead{Value} } 
\startdata
	$r_{\star}		$ &  $10^{15}$ cm 		\nl
	$z_{\star}		$ &  $2.0\times10^{15}$ cm 	\nl
	$\Gamma_{\star}		$ &  $2.0  $ 			\nl
	$\epsilon		$ &  $0.3  $ 			\nl
	$B_{\star}		$ &  $0.5 $ G 			\nl
	$m			$ &  $1.0  $		        \nl 
	$\Lambda_{kin}		$ &  $10^{45} $ ergs/s 		\nl
	$\gamma_{min}           $ &  $10^{2}	$		\nl
	$\gamma_{max}		$ &  $2.0\times 10^{5}  $       \nl
	$s			$ &  $2.0$    			\nl
	$\Theta			$ &  $10^{\circ}$

\enddata
\end{deluxetable}

\begin{deluxetable}{ccc}
\footnotesize
\tablecaption{Model parameters for PKS 2155--304. \label{tbl2}}
\tablewidth{0pt}
\tablehead{
\colhead{Parameter} & \colhead{Value} } 
\startdata
	$r_{\star}		$ &  $1.1\times10^{15}$ cm 		\nl
	$z_{\star}		$ &  $1.5\times10^{15}$ cm 		\nl
	$\Gamma_{\star}		$ &  $1.5  $ 				\nl
	$\epsilon		$ &  $0.3  $ 				\nl
	$B_{\star}		$ &  $0.11 $ G 				\nl
	$\Lambda_{kin}		$ &  $8.63\times 10^{45} $ ergs/s 	\nl
	$\gamma_{max}		$ &  $6.5\times 10^{5}  $      		\nl
	$s			$ &  $1.7$    				\nl
	$\Theta			$ &  $10^{\circ}$

\enddata
\end{deluxetable}

\begin{deluxetable}{ccc}
\footnotesize
\tablecaption{Values of parameters used in simulations under  the $\Theta$--$\Lambda$ scheme (cf. Fig. \ref{uni}).
 \label{tbl3}}
\tablewidth{0pt}
\tablehead{
\colhead{Parameter} & \colhead{Family A} & \colhead{Family B}  } 
\startdata

$r_{\star} $ &  $r$\tablenotemark{a}, $2^2 r$, $2^4 r$ &  $r$\tablenotemark{a}, $2^2 r$, $2^4 r$		\nl
$z_{\star} $ &  $z$\tablenotemark{b}, $2^2 z$, $2^4 z$ &  $z$\tablenotemark{b}, $2^2 z$, $2^4 z$		\nl
$\Gamma_{\star}	$ &  $1.5  $ &  $2.0  $ 				\nl
	$\epsilon$ &  $0.3  $ 	&  $0.4  $ 				\nl
	$B_{\star}$ &  $0.4 $ G &  $0.4 $ G     			\nl
	$\Lambda_{kin}$& $\Lambda$\tablenotemark{c}, $2^4 \Lambda$, $2^8 \Lambda$  &  $\Lambda$\tablenotemark{c}, $2^4 \Lambda$, $2^8 \Lambda$ 					       \nl
$\gamma_{max}$  & $4\times 10^{5}  $  &  $8\times 10^{5}  $    		\nl
	$s			$ &  $2.0$    			&  $1.7$

\tablenotetext{a}{$r=2.5 \times 10^{14}$ cm}
\tablenotetext{b}{$z=5.0 \times 10^{14}$ cm}
\tablenotetext{c}{$\Lambda=10^{43}$ ergs/s}

\enddata
\end{deluxetable}

\end{document}